\documentclass[a4paper,11pt]{article}
\pdfoutput=1 % if your are submitting a pdflatex (i.e. if you have
\usepackage{jcappub} % for details on the use of the package, please
\usepackage[T1]{fontenc} % if needed

\title{Structure formation in the new Deser-Woodard nonlocal gravity model}

\author[a]{Jia-Cheng Ding,}
\author[a,1]{Jian-Bo Deng,\note{Corresponding author.}}

\affiliation[a]{Institute of Theoretical Physics, Lanzhou University, Lanzhou 730000, P. R. China}
\emailAdd{dingjch16@lzu.edu.cn}
\emailAdd{dengjb@lzu.edu.cn}

\abstract{We consider the structure formation in the nonlocal gravity model proposed recently by Deser and Woodard (DW-2019 model), which can not only reproduce the $\Lambda$CDM cosmology without fine-tuning puzzle but also may provide a screening mechanism for free. By using the direct numerical method of the reconstructing technique, the nonlocal distortion function $f(Y)$ is fixed as $f(Y)\simeq\boldsymbol{e}^{2.153(Y-16.97)}$ which has a small deviation with the fitted function proposed by Deser and Woodard. Based on the numerical results, we plotted the curve of the growth rate $f\sigma_8(z)$ under the DW-2019 model, which shows this model is not ruled out by the $f\sigma_8$ data from the Redshift-space distortions measurements. The evolving curve of the growth rate has a distinct plummet at $z\simeq0.39$, which is a great difference with the DW model. The possible reason is that the DW-2019 model provides the strong nonlocal effect, which behaves as the anisotropic effect to influence the matter perturbation in the low-redshift range. Finally the qualitative analysis of the screening mechanism is discussed by considering the spacial dependence of the nonlocal modification in the small-scale range.}


\begin{document}
\maketitle
\flushbottom

\section{Introduction}
\label{sec:introduction}

Since the late-time accelerated expansion of universe was first detected 20 years ago~\cite{riess1998observational,perlmutter1999measurements}, it has aroused great interest among physicists. But the physics behind it is still under debate. Although Einstein¡¯s gravitational field equations are in remarkable agreement with all solar-system and binary-pulsar tests~\cite{will2006confrontation}, it cannot give a reasonable solution of the accelerated expansion of current universe, which suggests that Einstein gravity may not be applicable on the cosmological scale. Theoretically, the methods to produce an accelerated expansion of universe can be divided into two categories. The first one is to introduce the extra and assumptive component of matter, called dark energy, without changing the geometric terms of Einstein field equation. The other is to modify Einstein-Hilbert action to provide extra geometric terms in field equation. The $\Lambda$ cold dark matter ($\Lambda$CDM) model, belonging to the first category, can explain the late-time accelerated expansion of the universe well, and the influence of dark energy in the form of a cosmological constant $\Lambda$ is interpreted as the energy density of the vacuum. However, this otherwise formally and observationally consistent model carries two unsolved puzzles: the so-called coincidence and the fine-tuning problems. The former issue is that $\Lambda$CDM can not explain why the accelerated phase in the expansion began only recently in the cosmological time, while the latter expresses the enormous disagreement between the energy scale introduced by $\Lambda$ and the predictions of the standard model of particle physics for the vacuum energy density. Despite these two puzzles, $\Lambda$CDM model is still regarded as the standard model in astronomy for its simplicity of structure. On the other hand, many modified gravities belonging to the second category were proposed continually, including two major theories: the scalar-tensor theory~\cite{ratra1988cosmological,wetterich1988cosmology} and $f(R)$ theories~\cite{tsamis1998nonperturbative,capozziello2006dark,nojiri2007introduction,woodard2007avoiding}. In order to fit observed data, these modified models are required to emulate the background expansion history of the universe given by $\Lambda$CDM model via the reconstruction process~\cite{esposito2001scalar,saini2000reconstructing}. Then one can observationally distinguish among models by looking at their predictions beyond the background, such as solar system tests and the structure formation in the universe. However, these modified gravities still not avoid the fine-tuning puzzle.
\par
  Recently, a new type of modified gravities, nonlocal gravity, has aroused great interest because it can avoid fine-tuning successfully. The first nonlocal gravity model was proposed by Wetterich~\cite{wetterich1998effective}, who considered the following action
 \begin{equation}
 \label{the action of Wetterich}
 \mathcal{L}^{(W)}_{nonlocal}=\frac{1}{16\pi G}\big(R-g^{2}R\,\Box^{-1}R\big)\sqrt{-g},
 \end{equation}
 where $g^{2}$ is a dimensionless constant. $\Box^{-1}R$ is the inverse d'Alembertian acting on the Ricci scalar and it represents current effects of the necessarily abundant infrared gravitons in the early universe~\cite{woodard2014nonlocal,woodard2018case}. In the radiation-dominated era ($R=0$), the nonlocal term ``$R\Box^{-1}R$'' vanishes until universe enters into matter-dominated era. Hence, nonlocal model can naturally incorporate a delayed response to the transition from radiation to matter dominated era, yet avoid major fine-tuning. Unfortunately, the Wetterich model can not produce a viable cosmological evolution~\cite{wetterich1998effective}. Subsequently, other forms of nonlocal modified term were put forward consecutively, such as ``$m^2\Box^{-1}R$''~\cite{vardanyan2018nonlocally,Vardanyan_2018}, ``$R\Box^{-2}R$''~\cite{maggiore2014nonlocal,codello2018unified,PhysRevD.94.043531}, ``$R_{\mu\nu}\Box^{-1}R_{\mu\nu}$''~\cite{nersisyan2017instabilities,tian2018scalar}, ``$\mathcal{G}\Box^{-1}\mathcal{G}$''~\cite{capozziello2009accelerating} where $\mathcal{G}$ is the Gauss-Bonnet invariant ($\mathcal{G}\equiv R^{2}-4R_{\mu\nu}R^{\mu\nu}+R_{\mu\nu\rho\sigma}R^{\mu\nu\rho\sigma}$). Although these different forms of nonlocal term may produce a viable solution of the accelerated expansion of current universe to some extent, they have lost the structural simplicity.
\par
 In 2007, Deser and Woodard proposed a concise general form from the Wetterich model~\cite{PhysRevLett.99.111301}, called DW model. The action of this model is written as
 \begin{equation}
 \label{the action of DW}
 \mathcal{L}^{(DW)}_{nonlocal}=\frac{1}{16\pi G}R\big[1+f(X[g])\big]\sqrt{-g},
 \end{equation}
 where $X[g]\equiv \Box^{-1}R$ is dimensionless which is the same as in the Wetterich model. After the generalization of ``$X$'' to ``$f(X)$'', DW model obtains more freedom to simulate the $\Lambda$CDM cosmology without losing the simplicity of structure. After the reconstructing process, the nonlocal distortion function $f(X)$ is fixed as~\cite{deffayet2009reconstructing}
 \begin{equation}
 \label{the action of DW theory}
 \begin{split}
 f(X)\simeq0.245\big\{&\tanh\big[0.35(X+16.5)+0.032(X+16.5)^2+0.003(X+16.5)^3\big]-1\big\}.
 \end{split}
 \end{equation}
 In order to verify its reasonability, in \cite{nersisyan2017structure} authors studied the growth rate $f \sigma_{8}$ predicted by the DW model in $\Lambda$CDM background, and found that this model leads to a good agreement with the Redshift-space distortions observations (RSD) data as shown in FIG.\ref{f_sigma8}. RSD observations, which is one of the important tools in cosmology, can provide the information regarding the velocity field, probe the dark energy and test the gravity on the cosmological scale. A series of estimations for the cosmic growth rate at different redshift have been constrained by the RSD models, and provide a big database for testing the gravity.
\par
 However, DW model~\eqref{the action of DW theory} still has a ineluctable question. In \cite{deffayet2009reconstructing}, authors assumed that $X$ had opposite signs, in the cosmological $(-)$ and the (smaller scale) gravitationally bound $(+)$ contexts, which may provide a free screening mechanism. However, Ref.\cite{belgacem2019testing} pointed out that $X$ was negative definite without the expected screening mechanism, which contradicted the assumption in \cite{deffayet2009reconstructing}. In order to avoid this question, Deser and Woodard proposed a new nonlocal model~\cite{deser2019nonlocal}, called DW-2019 model. Its action is written as
 \begin{equation}
 \label{the action of DW2019}
 \mathcal{L}_{nonlocal}=\frac{\sqrt{-g}}{16\pi G}
  \left(
    \begin{array}{ccc}
    R+R f(Y[g])
    \end{array}
  \right),
 \end{equation}
 and the nonlocal scalar $Y[g]$ is given by
 \begin{equation}
 \label{Y}
 Y[g]\equiv\Box^{-1}(g^{\mu\nu}\partial_{\mu}X[g]\partial_{\nu}X[g]),
 \end{equation}
 where $\Box^{-1}$ is defined by retard boundary conditions which requires that $X$, $Y$ and their first derivatives all vanish on the initial value surface. As shown in \cite{deser2019nonlocal}, without losing the explanation of accelerated expansion, $Y$ has opposite signs in strongly bound matter $(-)$ and in the large-scale $(+)$ spontaneously. In the meantime, $Y$ still vanishes during radiation-dominated era just as $X$, and only grow slowly from then on. After the reconstructing process, the fitted nonlocal distortion function $f(Y)$ of DW-2019 model is proposed in \cite{deser2019nonlocal},
 \begin{equation}
 \label{f(Y)_DW-2019}
 f(Y)\simeq\boldsymbol{e}^{1.1(Y-16.7)}.
 \end{equation}
\par
 In this paper, we will verify the self-consistency and reasonability of the reconstructed DW-2019 model via the effective dark energy analysis as well as the fitting with RSD measurements. In order to calculate more accurately, firstly we will reconstruct DW-2019 model to obtain the numerical results which can simulate the $\Lambda$CDM background. And these numerical results will be used to calculate the structural growth rate of universe. Then we discuss the free screening mechanism of DW-2019 model qualitatively.
\par
 The rest of the paper is organized as follows: Sec.~\ref{sec:The Model} reviews DW-2019 model~\cite{deser2019nonlocal} and perturbs the model around the background solution to obtain the first-order perturbed equations that govern the growth of structure. In Sec.~\ref{sec:Numerical Analysis}, by applying the numerical method we reconstructed DW-2019 model to simulate the $\Lambda$CDM cosmology and analysed the evolution of the growth rate $f\sigma_{8}$. In Sec.~\ref{sec:Gravitational Slip Equation}, we discussed the gravitational slip to estimate the anisotropic effect from the nonlocal contribution in the DW-2019 model. Sec.~\ref{sec:The Free Screening Mechanism} shows a possible screening mechanism provided by the DW-2019 model. The last section is conclusions.

\section{The Model}
\label{sec:The Model}

 In this section, we review the background equations and derive the first-order perturbed equations provided by the DW-2019 model~\cite{deser2019nonlocal}.
\par
 Via the introduction of four auxiliary scalar fields ($X$,$Y$,$V$,$U$), the nonlocal version~\eqref{the action of DW2019} is localized as
 \begin{equation}
 \begin{aligned}
 \label{the action of local-DW2019}
 \mathcal{L}_{local}=\frac{\sqrt{-g}}{16\pi G}
  \big[R(1+U+f(Y))+g^{\mu\nu}(\partial_{\mu}X\partial_{\nu}U+\partial_{\mu}Y\partial_{\nu}V+V\partial_{\mu}X\partial_{\nu}X)\big].
 \end{aligned}
 \end{equation}
 Variation with respect to the auxiliary scalars respectively leads to the scalar equations
 \begin{align}
 \label{Box X}
 &\Box X=R,\\
 \label{Box Y}
 &\Box Y=g^{\mu\nu}\partial_{\mu}X\partial_{\nu}X,\\
 \label{Box V}
 &\Box V=R\,f^{(1)}(Y),\\
 \label{Box U}
 &\Box U=-2\nabla_{\mu}(V \nabla^{\mu}X),
 \end{align}
 where $\nabla_{\mu}$ is the covariant derivative operator compatible with $g_{\mu\nu}$ and $f^{(n)}(Y)$ is the n-order derivative of $f(Y)$ with respect to $Y$.
\par
 Variation of Eq.~\eqref{the action of local-DW2019} with respect to metric $g_{\mu\nu}$ yields the modified gravitational field equations,
 \begin{equation}
 \label{Deser_2019-field equation}
 G_{\mu\nu}+\Delta G_{\mu\nu}=8\pi G T_{\mu\nu},
 \end{equation}
 where $\Delta G_{\mu\nu}$ is the nonlocal modification, defined by
 \begin{equation}
 \begin{aligned}
 \label{Delta G_uv}
 \Delta G_{\mu\nu}\equiv&(G_{\mu\nu}+g_{\mu\nu}\Box-\nabla_{\mu}\nabla_{\nu})(U+f(Y))\\
 &+\partial_{(\mu}X\partial_{\nu)}U+\partial_{(\mu}Y\partial_{\nu)}V+V\partial_{\mu}X\partial_{\nu}X\\
 &\quad-\frac{1}{2}g_{\mu\nu}(\partial^{\alpha}X\partial_{\alpha}U+\partial^{\alpha}Y\partial_{\alpha}V+V\partial^{\alpha}X\partial_{\alpha}X).
 \end{aligned}
 \end{equation}
 It is evident that the nonlocal modification is covariantly conserved ($\nabla^{\mu}\Delta G_{\mu\nu}=0$), since it has been derived from a diff-invariant action. So the energy-momentum conservation $\nabla^{\mu}T_{\mu\nu}$ holds.

\subsection{The background equations}
\label{subsec:The background equations}

 Because of the homogeneity and isotropy of universe, it is worthy mentioning that the background is independent of spatial position, which leads that the background auxiliary scalars $\bar{X}$, $\bar{Y}$, $\bar{V}$, $\bar{U}$ (the bar denotes the background term) are only the time-dependent functions. Based on Friedman-Lema\^{\i}tre-Robertson-Walker (FLRW) metric in the conformal time $\tau$ $\big( d\tau\equiv\frac{1}{a}dt \big)$under the $(+,-,-,-)$ convention
 \begin{equation}
 \label{FLRW metric}
 ds^{2}=a(\tau)^2 \left[d\tau^2- d\boldsymbol{x}\cdot d\boldsymbol{x}\right],
 \end{equation}
 the $(00)$ and $(11)$ components of field equations are respectively given by
 \begin{equation}
 \begin{aligned}
 \label{00-component conformal time}
 3\mathcal{H}^2+\Delta \bar{G}_{00}=&8\pi G a^{2}\bar{\rho},\\
 \indent&\\
 \Delta \bar{G}_{00}=&3\mathcal{H}^2\left(\bar{U}+f(\bar{Y})\right)\\
 &+3\mathcal{H}\left(\bar{U}'+f^{(1)}(\bar{Y})\bar{Y}'\right)\\
 &+\frac{1}{2}\left(\bar{X}'\bar{U}'+\bar{Y}'\bar{V}'+\bar{V}\bar{X}'^2\right),
 \end{aligned}
 \end{equation}
 \begin{equation}
 \begin{aligned}
 \label{11-component conformal time}
 2\mathcal{H}'+\mathcal{H}^2-&\Delta \bar{G}_{11}=-8\pi G a^{2}\bar{p},\\
 \indent&\\
 \Delta \bar{G}_{11}=&-\left(2\mathcal{H}'+\mathcal{H}^2\right)\left(\bar{U}+f(\bar{Y})\right)\\
 &-\left[\bar{U}''+f^{(2)}(\bar{Y})\bar{Y}'^2+f^{(1)}(\bar{Y})\bar{Y}''\right]\\
 &-\mathcal{H}\left(\bar{U}'+f^{(1)}(\bar{Y})\bar{Y}'\right)\\
 &+\frac{1}{2}\left(\bar{X}'\bar{U}'+\bar{Y}'\bar{V}'+\bar{V}\bar{X}'^2\right),
 \end{aligned}
 \end{equation}
 where the prime denotes differentiation with respect to the conformal time $\tau$ and $\mathcal{H}\equiv\frac{a'}{a}$. $\bar{\rho}$ and $\bar{p}$ are the energy density and pressure without dark energy. The background scalar equations are
 \begin{align}
 \label{X}
 &\bar{X}''+2\mathcal{H}\bar{X}'=-6(\mathcal{H}'+\mathcal{H}^{2}),\\
 \label{Y}
 &\bar{Y}''+2\mathcal{H}\bar{Y}'=\bar{X}'^2,\\
 \label{V}
 &\bar{V}''+2\mathcal{H}\bar{V}'=-6(\mathcal{H}'+\mathcal{H}^{2})f^{(1)}(\bar{Y}),\\
 \label{U}
 &\bar{U}''+2\mathcal{H}\bar{U}'=-2\bar{X}'\bar{V}'+12\bar{V}(\mathcal{H}'+\mathcal{H}^{2}).
 \end{align}
 These background equations will be used later.

 \subsection{The first-order perturbed equations}
 \label{subsec:The first-order perturbed equations}

 In this section, we discuss the linear scalar perturbation equations for DW-2019 model and the method we used is similar to one implemented in \cite{nersisyan2017structure,PhysRevD.90.043535}. We use the same symbol convention as in \cite{baumann2015cosmology}.
\par
 Firstly, we introduce the perturbed metric under the Newtonian gauge
 \begin{equation}
 \label{the perturbed metric of g_uv}
 g_{\mu\nu}=a(\tau)^2
    \left[
    \begin{array}{ccc}
    1+2\Psi(\tau, \boldsymbol{x})& 0  \\
    0 & -(1-2\Phi(\tau,\boldsymbol{x}))\delta_{ij} \\
    \end{array}
    \right].
 \end{equation}
 The perturbed scalar auxiliary fields can be decomposed into the background term and the perturbation,
 \begin{equation}
 \label{F=X-Y-V-U}
 J(\tau,\boldsymbol{x})=\bar{J}(\tau)+\delta J(\tau, \boldsymbol{x})\indent(J=X,Y,V,U).
 \end{equation}
 The d'Alembertian acting on $J$ is expanded as
 \begin{equation}
 \begin{aligned}
 \label{the scalar d'Alembertian-J}
 \Box J=\frac{1}{a^2}\big\{(1-2\Psi)\bar{J}''+\big[2\mathcal{H}(1-2\Psi)-(\Psi'+3\Phi')\big]\bar{J}'+\delta J''+2\mathcal{H}\mathcal\delta J'-\nabla^{2}\delta J\big\},
 \end{aligned}
 \end{equation}
 where we used $\Box J=g^{\alpha\beta}(\partial_{\alpha}\partial_{\beta}J-\Gamma^{\lambda}_{\alpha\beta}\partial_{\lambda}J)$, $\Gamma^{\lambda}_{\alpha\beta}$ is the Christoffel symbol compatible with the perturbed metric.
\par
 The first-order equations of the perturbed scalar equations are
 \begin{equation}
 \begin{aligned}
 \label{the first-order equation of Box-X}
 \delta X''+2\mathcal{H}\delta X'\!-\!\boldsymbol{\nabla}^{2}\delta X-(\Psi'+3\Phi')\bar{X}'-6\Phi''-6\mathcal{H}\,(\Psi'+3\Phi')-2\boldsymbol{\nabla}^{2}(\Psi-2\Phi)=0,\indent\indent
 \end{aligned}
 \end{equation}
 \begin{equation}
 \begin{aligned}
 \label{the first-order equation of Box-Y}
 \delta Y''+2\mathcal{H}\delta Y'-&\boldsymbol{\nabla}^{2}\delta Y-(\Psi'+3\Phi')\bar{Y}'-2\bar{X}'\delta X'=0,\indent\indent\indent\indent\indent\indent\indent\indent\indent\indent\indent
 \end{aligned}
 \end{equation}
 \begin{equation}
 \begin{aligned}
 \label{the first-order equation of Box-V}
 \delta V''+2\mathcal{H}\delta V'-&\boldsymbol{\nabla}^{2}\delta V-(\Psi'+3\Phi')\bar{V}'-6\Phi''\,f^{(1)}(\bar{Y})\indent\indent\\
 &\indent\indent\qquad-6\mathcal{H}\,f^{(1)}(\bar{Y})\,(\Psi'+3\Phi')-2\boldsymbol{\nabla}^{2}(\Psi-2\Phi)\,f^{(1)}(\bar{Y})=0,\indent\indent
 \end{aligned}
 \end{equation}
 \begin{equation}
 \begin{aligned}
 \label{the first-order equation of Box-U}
 \delta U''+&2\mathcal{H}\delta U'-\boldsymbol{\nabla}^{2}\delta U-(\Psi'+3\Phi')\bar{U}'+12\bar{V}\Phi''\indent\indent\indent\indent\indent\indent\indent\indent\indent\indent\\
 &\indent\indent\indent+12\mathcal{H}\,\bar{V}\,(\Psi'+3\Phi')-12\,\delta V\,(\mathcal{H}'+\mathcal{H}^{2})\\
 &\indent\indent\indent\indent\indent\indent\quad+2\,(\bar{X}'\delta V'+\delta X'\bar{V}')+4\bar{V}\,\boldsymbol{\nabla}^{2}(\Psi-2\Phi)=0,
 \end{aligned}
 \end{equation}
 where $\boldsymbol{\nabla}^{2}$ is the Laplacian operator and $f^{(1)}(Y)\simeq f^{(1)}(\bar{Y})$. The metric perturbation fields can be composed into spatial plane waves
 \begin{equation}
 \begin{aligned}
 \label{the spatial plane waves}
 &\Psi(\tau, \boldsymbol{x})\equiv\int\frac{d^{3}k}{(2\pi)^{3}}\boldsymbol{e}^{i\boldsymbol{k}\cdot\boldsymbol{x}}\Psi(\tau, \boldsymbol{k}),\\
 &\Phi(\tau, \boldsymbol{x})\equiv\int\frac{d^{3}k}{(2\pi)^{3}}\boldsymbol{e}^{i\boldsymbol{k}\cdot\boldsymbol{x}}\Phi(\tau, \boldsymbol{k}).
 \end{aligned}
 \end{equation}
 In Fourier space, considering the sub-horizon limit ($k\gg \mathcal{H}$), Eqs.\eqref{the first-order equation of Box-X}-\eqref{the first-order equation of Box-U} gives
 \begin{align}
 \label{delta X}
 &\delta X=-(2\Psi-4\Phi),\\
 \label{delta Y}
 &\delta Y\simeq0,\\
 \label{delta V}
 &\delta V=-(2\Psi-4\Phi)f^{(1)}(\bar{Y}),\\
 \label{delta U}
 &\delta U=4\bar{V}(\Psi-2\Phi).
 \end{align}
  Obviously, on the sub-horizon limit, for DW-2019 model, the first-order perturbation of the scalar field $Y$ vanishes, which may produce a discontinuous growth of the matter density perturbation as shown in FIG.\ref{delta-delta1}.
\par
 Generally, for the anisotropic fluid in the first-order perturbation, we have
 \begin{align}
 \label{T00}
 &T_{0}^{0}=\bar{\rho}+\delta\rho,\\
 \label{T0i}
 &T_{0}^{i}=(\bar{\rho}+\bar{p})v^{i},\\
 \label{Tij}
 &T_{j}^{i}=-(\bar{p}+\delta p)\delta_{j}^{i}-\Pi_{j}^{i},
 \end{align}
 where $v^{i}\equiv dx^{i}/d\tau$ is the coordinate velocity, $\Pi_{j}^{i}$ is the spatial part of the anisotropic stress tensor which is traceless.
 The first-order part of the ($00$) component of field equations is given by
 \begin{equation}
 \begin{aligned}
 \label{the frist-order part of 00-field equation}
 &2 \big[ \boldsymbol{\nabla}^{2}\Phi-3\mathcal{H}\Phi'-3\mathcal{H}^{2}\Psi \big] \big(1+\bar{U}+f(\bar{Y})\big)-\Psi \big( \bar{X}'\bar{U}'+\bar{Y}'\bar{V}'+\bar{V}\bar{X}'^{2} \big)\\
 &\indent-3 \big( \Phi'+2\mathcal{H}\Psi \big) \cdot \frac{\partial}{\partial\tau} \big( \bar{U}+f(\bar{Y}) \big) + \big[ 3\mathcal{H}^{2}+3\mathcal{H}\frac{\partial}{\partial\tau}-\boldsymbol{\nabla}^{2} \big] \big( \delta U+f^{(1)}(\bar{Y})\delta Y \big)\\
 &\indent\indent+\frac{1}{2} \big( \bar{X}'\delta U'+\delta X'\bar{U}'+\bar{Y}'\delta V'+\delta Y'\bar{V}'+2\bar{V}\bar{X}'\delta X'+\bar{X}'^{2}\delta V \big) =8\pi G a^{2}\delta\rho.
 \end{aligned}
 \end{equation}
 Considering the sub-horizon limit, it is reduced to
 \begin{equation}
 \begin{aligned}
 \label{the (00) component in the sub-horizon limt}
 \big(1+\bar{U}+f(\bar{Y})\big)\Phi-\frac{1}{2}\big(\delta U+f^{(1)}(\bar{Y})\delta Y\big)=-\frac{4\pi G a^{2}\delta\rho}{k^{2}}.
 \end{aligned}
 \end{equation}
 The first-order parts of the ($ij$) components of field equation is given by
 \begin{equation}
 \begin{aligned}
 \label{the frist-order part of ij-field equation}
 &\delta_{ij} \big[ \boldsymbol{\nabla}^{2}(\Psi-\Phi)+2\Phi''+2\Psi(2\mathcal{H}'+\mathcal{H}^{2})+2\mathcal{H}(\Psi'+2\Phi') \big] \big( 1+\bar{U}+f(\bar{Y}) \big)\\
 &-\delta_{ij}\Psi \big[ \bar{X}'\bar{U}'+\bar{Y}'\bar{V}'+\bar{V}\bar{X}'^{2} \big]-\delta_{ij} \big(2\mathcal{H}'+\mathcal{H}^{2}\big)\big(\delta U+f^{(1)}(\bar{Y})\delta Y\big)\\
 &+\delta_{ij}\bigg\{ \big[ 2\mathcal{H}(2\Phi+3\Psi)+(\Psi'+2\Phi') \big] \frac{\partial}{\partial\tau}+2\Psi\frac{\partial^{2}}{\partial\tau^{2}} \bigg\} \big( \bar{U}+f(\bar{Y}) \big)\\
 &+\delta_{ij}\big[3\mathcal{H}\frac{\partial}{\partial\tau}+\frac{\partial^{2}}{\partial\tau^{2}}-\boldsymbol{\nabla}^{2} \big]\big(\delta U+f^{(1)}(\bar{Y})\delta Y\big)\\
 &+\delta_{ij}\ \frac{1}{2} \big[ \bar{X}'\delta U'+\delta X'\bar{U}'+\bar{Y}'\delta V'+\delta Y'\bar{V}'+2\bar{V}\bar{X}'\delta X'+\bar{X}'^{2}\delta V \big]\\
 &+\big(1+\bar{U}+f(\bar{Y}) \big) \, \partial_{i}\partial_{j}(\Phi-\Psi)-\partial_{i}\partial_{j} \big( \delta U+f^{(1)}(\bar{Y})\delta Y \big)\\
 &\indent\indent\indent\indent\indent\indent\indent\indent\indent=8\pi G a^{2}\delta_{ij}\delta p-8\pi G a^{2}\Pi_{ij}.
 \end{aligned}
 \end{equation}
 Its trace-free parts are
 \begin{equation}
 \begin{aligned}
 \label{the trace-free part of (ij) component}
 \big(1+\bar{U}+&f(\bar{Y})\big)\partial_{i}\partial_{j}\big(\Phi-\Psi\big)-\partial_{i}\partial_{j}\big(\delta U+f^{(1)}(\bar{Y})\delta Y\big)=-8\pi G a^{2}\Pi_{ij}.
 \end{aligned}
 \end{equation}
 Without regard to the anisotropic stress, there is no source on the right-hand side, which leads to
 \begin{equation}
 \label{phi-psi}
 \Phi-\Psi=\frac{\delta U+f^{(1)}(\bar{Y})\delta Y}{1+\bar{U}+f(\bar{Y})}.
 \end{equation}
 In the sub-horizon limit, from Eqs.\eqref{delta Y}, \eqref{delta U}, \eqref{the (00) component in the sub-horizon limt} and \eqref{phi-psi}, the metric perturbations $\Psi$, $\Phi$ can be expressed in terms of the density perturbation
 \begin{equation}
 \begin{aligned}
 \label{solution of psi}
 &\Psi=-\frac {1+\bar{U}+f(\bar{Y})+8\,\bar{V}}{\left(1+\bar{U}+f(\bar{Y})\right)\left(1+\bar{U}+f(\bar{Y})+6\bar{V}\right)}\cdot\frac{4\pi G\, a^{2}\,\bar{\rho}\,\delta}{k^{2}}\\
 &\indent
 \end{aligned}
 \end{equation}
 \begin{equation}
 \begin{aligned}
 \label{solution of phi}
 &\Phi=-\frac {1+\bar{U}+f(\bar{Y})+4\,\bar{V}}{\left(1+\bar{U}+f(\bar{Y})\right)\left(1+\bar{U}+f(\bar{Y})+6\bar{V}\right)}\cdot\frac{4\pi G\, a^{2}\,\bar{\rho}\,\delta}{k^{2}},\\
 &\indent
 \end{aligned}
 \end{equation}
 where $\delta$ represents the fractional density perturbation ($\delta\equiv\frac{\delta \rho}{\rho}$).
\par
 In the matter dominated era, according to the perturbed energy-momentum conservation law~\cite{baumann2015cosmology}, one can get the equation for the matter density perturbation $\delta_{m}$ in the sub-horizon limit $\delta_{m}''+\mathcal{H}\delta_{m}'=-k^{2}\Psi$. Then we get the $k$-independent growth equation for the matter density perturbation $\delta_{m}$ in the DW-2019 model
 \begin{align}
 \label{the equation of delta m}
 \delta_{m}''+\mathcal{H}\delta_{m}'=G_{N}\cdot4\pi G a^{2}\bar{\rho}_{m}\,\delta_{m},
 \end{align}
 where
 \begin{equation}
 \begin{aligned}
 \label{GN}
 G_{N}\!\equiv\!\frac {1+\bar{U}+f(\bar{Y})+8\,\bar{V}}{\left(1+\bar{U}+f(\bar{Y})\right)\left(1+\bar{U}+f(\bar{Y})+6\bar{V}\right)}.
 \end{aligned}
 \end{equation}
 Then the metric perturbation $\Psi$ in the Fourier space can be written as
 \begin{equation}
 \begin{aligned}
 \label{GN}
 \Psi(\tau,k)=-\frac{4\pi G a^{2}\bar{\rho}_{m}}{k^{2}}\cdot G_{N}(\tau)\cdot\delta_{m}(\tau).
 \end{aligned}
 \end{equation}
 The above equations indicate that the matter perturbation $\delta_{m}$ generates the metric perturbation $\Psi$, in the meantime $\Psi$ influences the evolution of $\delta_{m}$.

\section{Numerical Analysis}
\label{sec:Numerical Analysis}

 As a trial, we used the fitted function of $f(Y)$\eqref{f(Y)_DW-2019} proposed in \cite{deser2019nonlocal} to test the EoS parameter of the effective dark energy component $w_{de}$. The result shows $w_{de}$ predicted by this fitted function is in contradiction with the precondition of the $\Lambda$CDM background ($w_{de}\sim-1$). This is because that the function $f(Y)$ is only an approximate function provided by the reconstructed numerical results and it does deviate the numerical result to some extent. In order to get the accurate calculations, we reconstructed DW-2019 model once again and the numerical result will be used to discuss the structure formation of DW-2019 model. Comparing with the reconstructing technique in \cite{deser2019nonlocal}, our reconstructing technique is more straightforward without complicated transformations.

\subsection{Specialization to $\Lambda$CDM}
\label{subsec:Specialization to Lambda CDM}
 For simplicity, the useful time variable $N=\ln a$ is always used. $N$ represents the number of $\boldsymbol{e}$-foldings until the present and the current scale factor $a_{0}$ is generally identified as $1$. Its various derivatives are
 \begin{equation}
 \begin{aligned}
 &\frac{d}{d\tau}=\boldsymbol{e}^{N}H\partial_{N},\\
 &\frac{d^2}{d \tau^2}=\boldsymbol{e}^{2N}H^{2}\big[\partial^{2}_{N}+(\xi+1)\,\partial_{N}\big].\\
 &\big(\,\xi\equiv\frac{1}{H}\partial_{N}H\,\big)
 \end{aligned}
 \end{equation}
  Based on the background scalar equations\eqref{X}-\eqref{U}, we get
  \begin{align}
  \label{X-N}
  &\partial^{2}_{N}\bar{X}+(\xi+3)\,\partial_{N}\bar{X}=-6(2+\xi),\\
  \label{Y-N}
  &\partial^{2}_{N}\bar{Y}+(\xi+3)\,\partial_{N}\bar{Y}=(\partial_{N}\bar{X})^2,\\
  \label{U-N}
  &\partial_{N}\bar{U}=-2\,\partial_{N}\bar{X}\,\bar{V}.
  \end{align}
\par
 For the purpose of emulating the $\Lambda$CDM cosmology, the form of Hubble parameter is chosen as
 \begin{equation}
 \label{Hubble parameter for the background}
 H=H_{0}\,\sqrt{\Omega_{r0}\,\boldsymbol{e}^{-4N}+\Omega_{m0}\,\boldsymbol{e}^{-3N}+\Omega_{\Lambda0}}
 \end{equation}
 where $\big(\Omega_{r0}, \, \Omega_{m0}\big)$ is fixed as $\big(9.265\!\times\!10^{-5}, \, 0.315\big)$ and the matter fluctuation amplitude $\sigma^{0}_{8}$ is fixed as $0.811$ based on Plank 2018~\cite{aghanim2018planck}. The symbol ``$\,0\,$'' denotes quantities evaluated today.
\par
 Generally, the initial conditions of scalar fields deep inside radiation dominated era $(N_{ini}=-16)$ are postulated as
 \begin{equation}
 \begin{aligned}
 \label{X-Y-V-U_initial conditio}
 &\bar{X}(N_{ini})=\partial_{N}\bar{X}(N_{ini})=0,\\
 &\bar{Y}(N_{ini})=\partial_{N}\bar{Y}(N_{ini})=0,\\
 &\bar{V}(N_{ini})=\partial_{N}\bar{V}(N_{ini})=0,\\
 &\bar{U}(N_{ini})=\partial_{N}\bar{U}(N_{ini})=0.
 \end{aligned}
 \end{equation}
 Actually, the initial conditions depend on the thermal history of the Universe as shown in \cite{PhysRevD.94.043531}, which points out that the nonzero initial conditions should not be ignored. In this paper, we do not focus on the situation with the nonzero initial conditions.
\par
 Based on the background equations of $\bar{X}$ in Eq.~\eqref{X-N} and $\bar{Y}$ in Eq.~\eqref{Y-N}, we can solve the equations of $\bar{X}$ and $\bar{Y}$ in numerical method. Furthermore, from the background field equations~\eqref{00-component conformal time} and \eqref{11-component conformal time}, one can get
 \begin{equation}
 \label{F-N}
 \partial^{2}_{N}\bar{F}+(\xi+5)\partial_{N}\bar{F}+(6+2\xi)\bar{F}=-\frac{6\Omega_{\Lambda0}}{h^{2}}
 \end{equation}
 where $\bar{F}\equiv\bar{U}+f(\bar{Y})$ and $h\equiv H/H_{0}$. We can obtain the numerical results of $\bar{F}$ from Eq.~\eqref{F-N} with the initial conditions
 \begin{equation}
 \begin{aligned}
 \label{the initil condition of F}
 \bar{F}(N_{ini})=0,\indent \partial_{N}\bar{F}(N_{ini})=0.
 \end{aligned}
 \end{equation}
 In order to obtain the solution of $\bar{U}$ and $\bar{V}$, we apply the method proposed in \cite{deser2019nonlocal}, defining $G\equiv-\partial_{N}\bar{U}/\partial_{N}\bar{X}$,
 \begin{equation}
 \label{G-N}
 (\partial_{N}+3+\xi)\partial_{N}G+12(2+\xi)\,\frac{\partial_{N}\bar{X}}{\partial_{N}\bar{Y}}\,G+12(2+\xi)\,\frac{\partial_{N}\bar{F}}{\partial_{N}\bar{X}}=0.
 \end{equation}
\par
\begin{figure}[tbp]
  \centering
  \includegraphics[width=.6\textwidth]{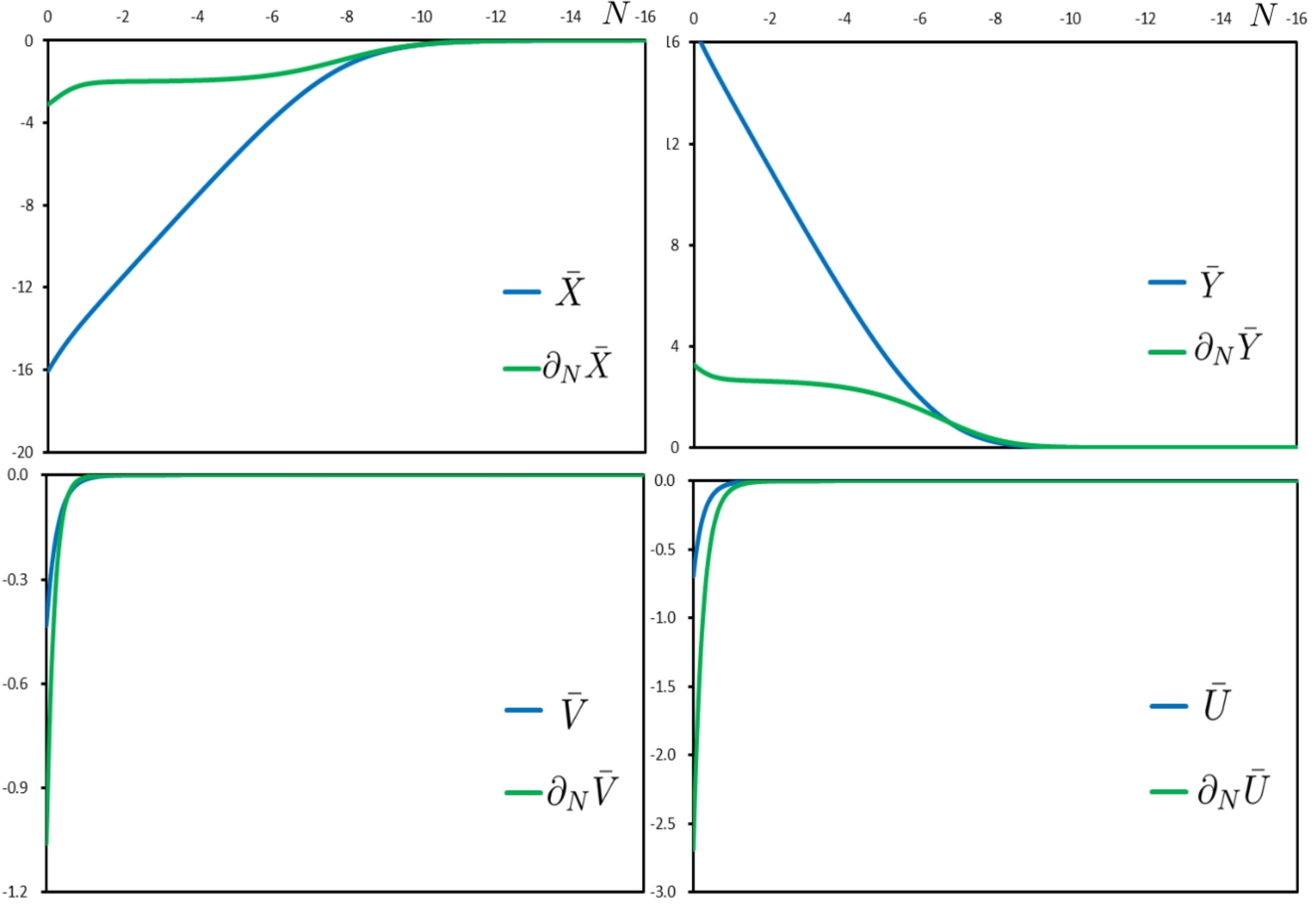}
  \caption{\label{X-Y-V-U} The evolution of the background scalars $\bar{X},\,\bar{Y},\,\bar{V},\,\bar{U}$ and their derivatives with respect to the $\boldsymbol{e}$-folding time $N$.}
 \end{figure}

 Based on the numerical results of $\bar{X}$, $\bar{Y}$ and $\bar{F}$, one can get the numerical result of $G$. It is worth mentioning that our numerical method is based on the fourth-order Runge-kutta method with discrete data. $\bar{V}$ and $\partial_{N}\bar{U}$ can be solved by $G=2\bar{V}$ and $\partial_{N}\bar{U}=-2\,\partial_{N}\bar{X}\,\bar{V}$, then one can get the numerical result of $\partial_N\bar{V}$ by substituting these results into Eq.\eqref{00-component conformal time}, shown in FIG.\ref{X-Y-V-U}.
\par
\begin{figure}[tbp]
  \centering
  \includegraphics[width=0.62\textwidth]{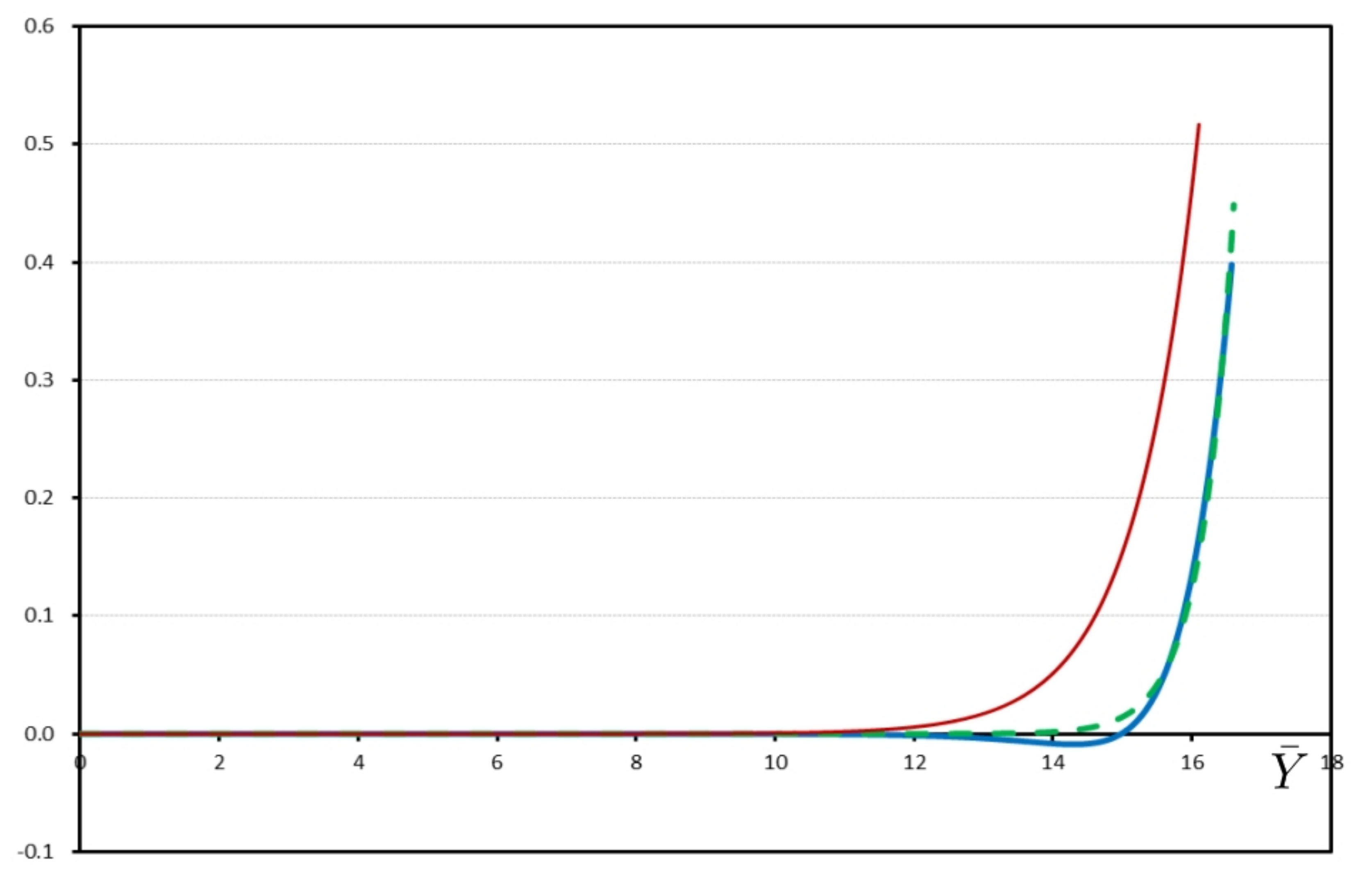}
  \caption{\label{f_Y_fit} The curves of the full numerical determination of $f(\bar{Y})$ (the solid blue curve), the resulting exponential fit \eqref{f(Y)_fit} (the dashed green curve) and the fitted function proposed by \cite{deser2019nonlocal} (the crimson curve).}
 \end{figure}

 Applying the numerical results and the one-to-one relation between $\bar{Y}(N)$ and $f(N)$, the nonlocal distortion function $f(\bar{Y})$ is fixed as
 \begin{equation}
 \begin{aligned}
 \label{f(Y)_fit}
 f(\bar{Y})\simeq\boldsymbol{e}^{\,2.153\,(\bar{Y}-16.97)}.
 \end{aligned}
 \end{equation}
 This fitted function has the small deviation with the result in \cite{deser2019nonlocal}, as shown in FIG.~\ref{f_Y_fit}, which may result from the small difference of the initial conditions.
\par
\begin{figure}[tbp]
 \centering
 \includegraphics[width=0.64\textwidth]{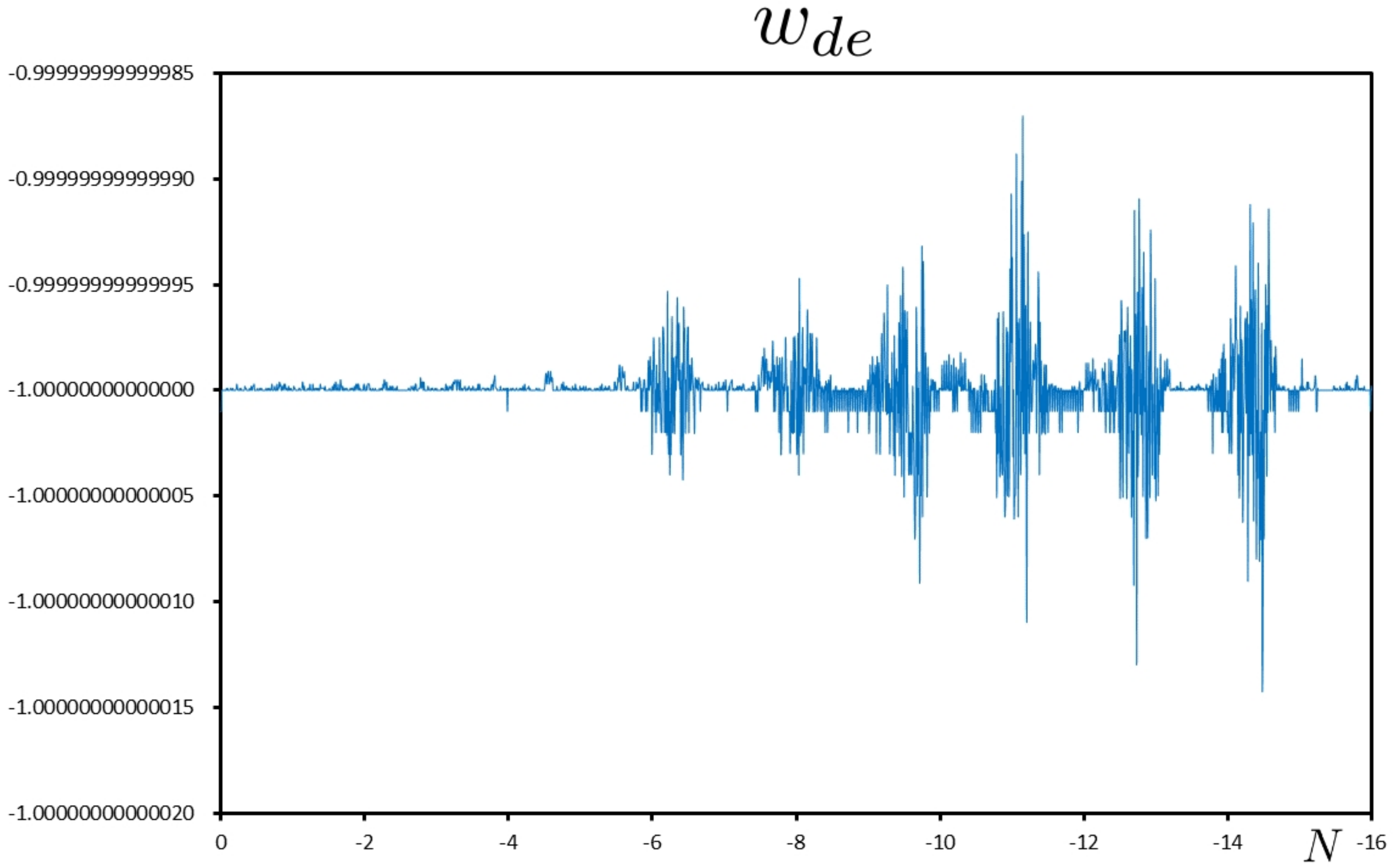}
 \caption{\label{the Eos parameter w(de) of dark enegy}The numerical result of the EoS parameter $w_{de}$ under the $\Lambda$CDM background in the DW-2019 model.}
 \end{figure}

 In order to test the self-consistency of our numerical results, we checked the EoS parameter $w_{de}$ of the effective dark energy component provided by the nonlocal modifications in DW-2019 model,
 \begin{equation}
 \begin{aligned}
 \label{the Eos parameter of dark energy density}
 w_{de}=\frac{\bar{p}_{de}}{\bar{\rho}_{de}},
 \end{aligned}
 \end{equation}
 where $\bar{\rho}_{de}\equiv-\frac{1}{8\pi G a^{2}}\Delta \bar{G}_{00}$ and $\bar{p}_{de}\equiv-\frac{1}{8\pi G a^{2}}\Delta \bar{G}_{11}$. The result in FIG.~\ref{the Eos parameter w(de) of dark enegy} shows $w_{de}$ approaches to $-1$ very closely, consisting with the precondition (the $\Lambda$CDM background) well.
\par

\subsection{The matter density perturbation in DW-2019 model}
\label{subsec:The matter density perturbation in DW-2019 model}

 In the $\boldsymbol{e}$-folding time $N$, the growth equation for the matter density perturbation~\eqref{the equation of delta m} can be transformed into
 \begin{equation}
 \begin{aligned}
 \label{delta m in N}
 \partial^{2}_{N}\delta_{m}+(\xi+2)\,\partial_{N}\delta_{m}=G_{N}\cdot\frac{3}{2}\,\Omega_{m0}\,\frac{H^{2}_{0}}{H^{2}}\,\boldsymbol{e}^{-3N}\,\delta_{m}.
 \end{aligned}
 \end{equation}
 The initial conditions of $\delta_{m}$ deep into the matter dominated era are taken to consist with the pure CDM model~\cite{nersisyan2017structure}
 \begin{equation}
 \label{delta m-initial condition}
 \delta_{m}(N^{*}_{ini})=a^{*}_{ini},\indent\frac{\partial_{N}\delta_{m}(N^{*}_{ini})}{\delta_{m}(N^{*}_{ini})}=1,\indent(N^{*}_{ini}=\ln{a^{*}_{ini}})
 \end{equation}
 where the initial scalar factor $a^{*}_{ini}$ is taken at redshift $z^{*}_{ini}=9$.
\par
\begin{figure}[tbp]
 \centering
 \includegraphics[width=0.6\textwidth]{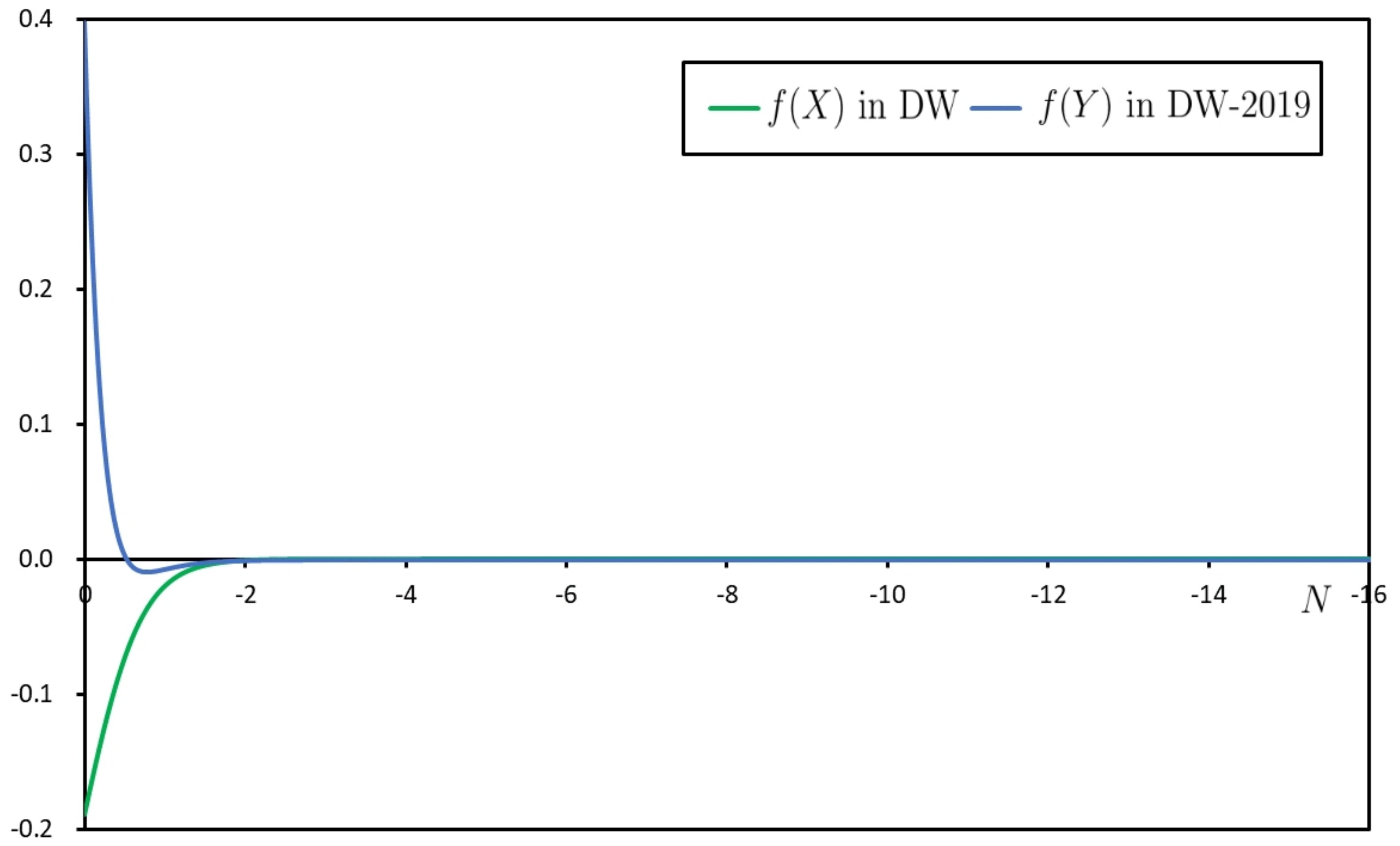}
 \caption{\label{The nonlocal distortion functions in the DW and DW-2019 models} The comparison of the nonlocal distortion function in the DW model and the DW-2019 model via numerical methods.}
 \end{figure}
\begin{figure}[tbp]
 \centering
 \includegraphics[width=0.6\textwidth]{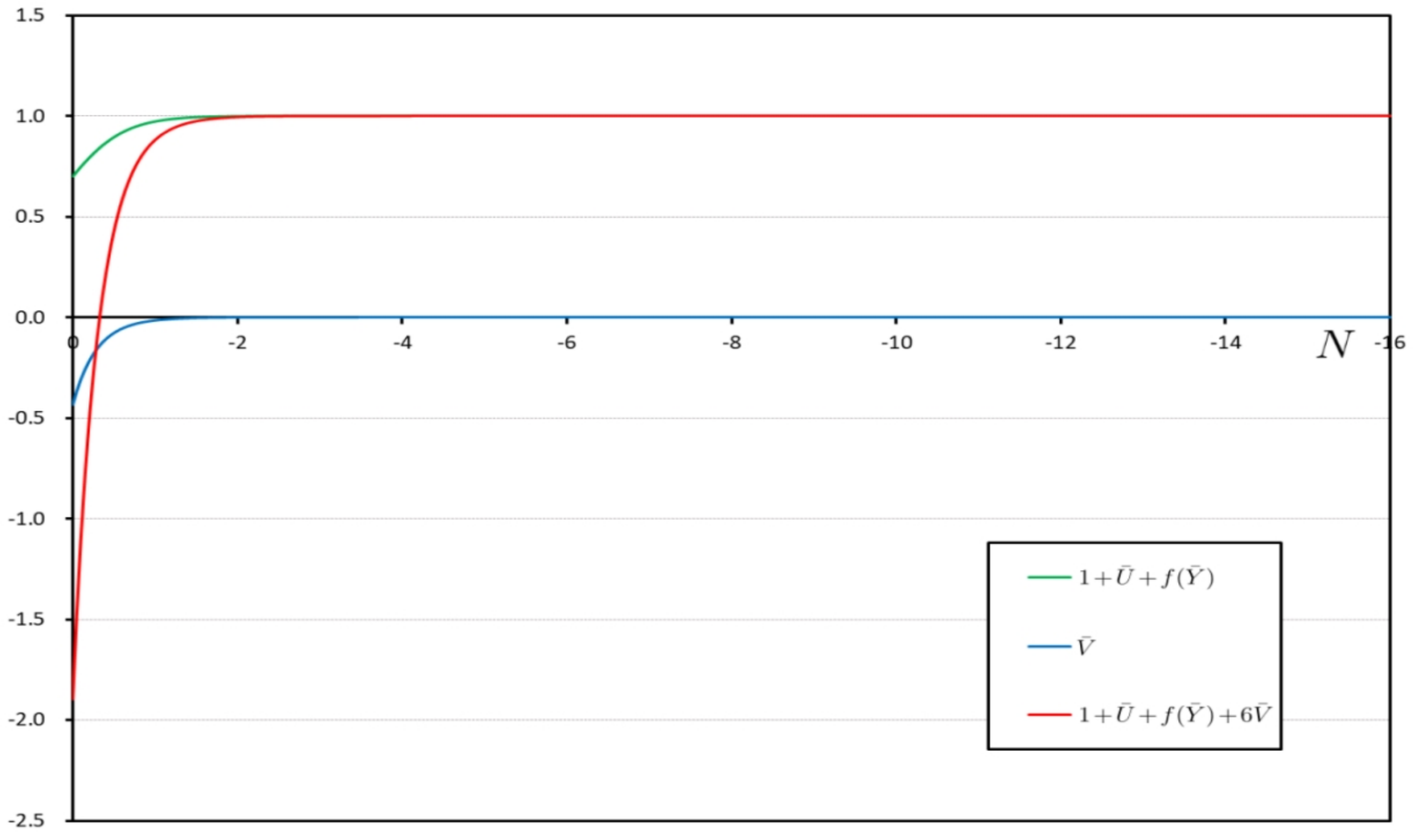}
 \caption{\label{1_f_U and V} The numerical results of $(1+\bar{U}+f(\bar{Y}))$, $\bar{V}$ and $(1+\bar{U}+f(\bar{Y})+6\bar{V})$.}
 \end{figure}
\begin{figure}[tbp]
 \centering
 \includegraphics[width=0.6\textwidth]{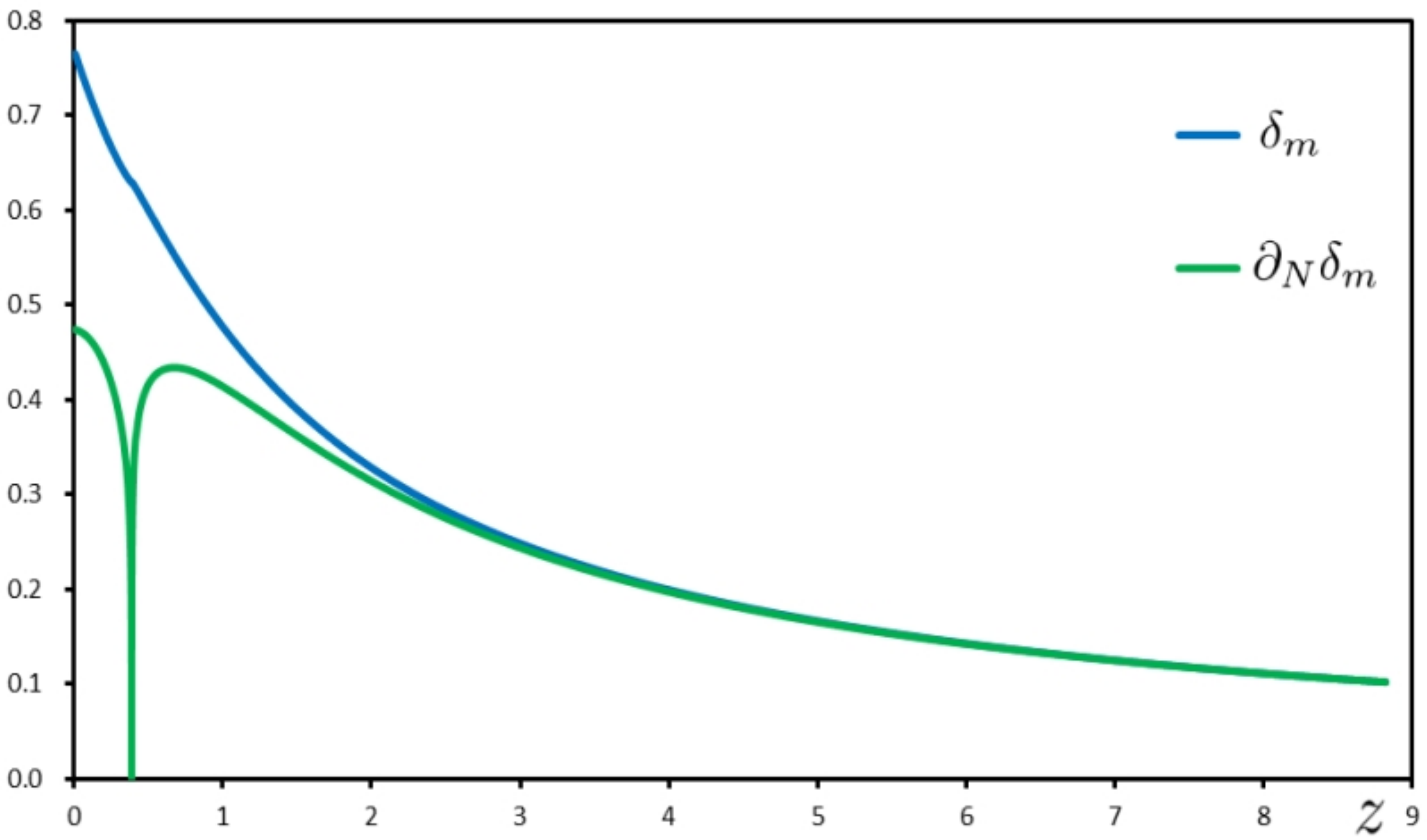}
 \caption{\label{delta-delta1} The matter perturbation $\delta_{m}$ and its derivative $\partial_{N}\delta_{m}$ at different redshift $z$ in the DW-2019 model under the $\Lambda$CDM background\eqref{Hubble parameter for the background}.}
 \end{figure}

 Based on the numerical results from reconstructing process, we solved Eq.~\eqref{delta m in N} numerically. As shown in FIG.~\ref{delta-delta1}, the evolving curve of $\partial_{N}\delta_{m}$ has a peculiar null point at $z\simeq0.39$ that leads to the discontinuity of $\delta_{m}$, which is different from the case of the DW model. In order to explain this puzzle, we focused on the difference between the DW model and the DW-2019 model. The nonlocal distortion function of the DW model is defined as the function of the scalar field $\bar{X}$ that sources from the Ricci scalar of the universe as shown in Eq.\eqref{Box X}. For the DW-2019 model, the nonlocal distortion function is defined as the function of the new scalar field $\bar{Y}$ whose source is proportional to the square of the rate of change of $\bar{X}$ as shown in Eq.\eqref{Box Y}. The evolving curves of these two nonlocal distortion functions are illustrated in FIG.~\ref{The nonlocal distortion functions in the DW and DW-2019 models}, which shows the amplitude of variation of $f(\bar{Y})$ of the DW-2019 model is greater than $f(\bar{X})$ of the DW model. Hence the form of nonlocal modification in the action of the DW-2019 model may provide a stronger nonlocal effect than that of the DW model. On the other hand, based on the numerical results as illustrated in FIG.~\ref{1_f_U and V}, beyond the $\Lambda$CDM background, we find that the auxiliary scalar field $\bar{V}$ is negative and the term of $(1+\bar{U}+f(\bar{Y}))$ keeps positive, and $\bar{V}$ decreases faster than the increasing rate of $(1+\bar{U}+f(\bar{Y}))$, which makes $1+\bar{U}+f(\bar{Y})+6\bar{V}\rightarrow0$ at $N\simeq-0.328$ ($z\simeq0.39$) so that $G_{N}$ is divergent at this point. These numerical characteristics directly lead to the discontinuities of $\partial_{N}\delta_{m}$ and the growth rate $f\sigma_{8}$.
\par
 In brief, the nonlocal effect behaves as a cumulative effect. In the high-redshift range this cumulative effect is weak and the DW-2019 model can reduce to GR. After a long time with the accumulation and turning into the low-redshift range, the nonlocal effect is stronger enough to impact the matter perturbation.
\par
 The measurements of the growth rate $f\sigma_{8}$ at different redshift $z$ can be used to test the theories of dark energy and the modified gravities. $f$ represents the structural growth rate of universe and $\sigma_{8}$ is the amplitude of matter fluctuations in spheres of $8\,h^{-1}$ Mpc, defined by
 \begin{equation}
 \begin{aligned}
 \label{f and sigma8}
 f\equiv\partial_{N}(\ln\delta_{m}),\quad
 \sigma_{8}(N)\equiv\sigma^{0}_{8}\,\frac{\delta_{m}(N)}{\delta_{m}(0)}.
 \end{aligned}
 \end{equation}

 \begin{figure}[tbp]
 \centering
 \includegraphics[width=0.6\textwidth]{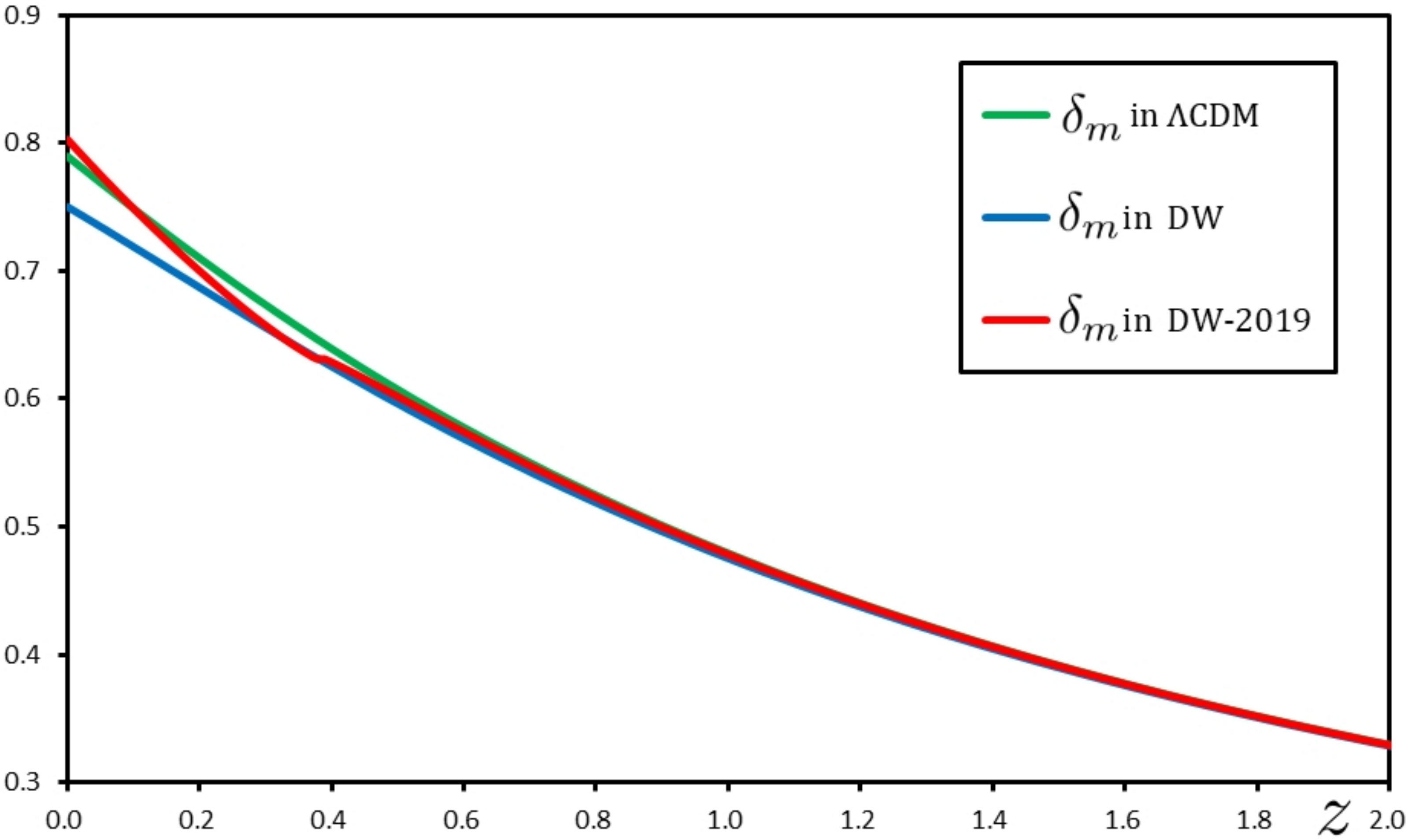}
 \caption{\label{delta_compare} The comparison of the different matter perturbation $\delta_{m}(z)$ in the $\Lambda$CDM, DW model and DW-2019 model.}
 \end{figure}

\begin{figure}[tbp]
\centering
\includegraphics[width=0.6\textwidth]{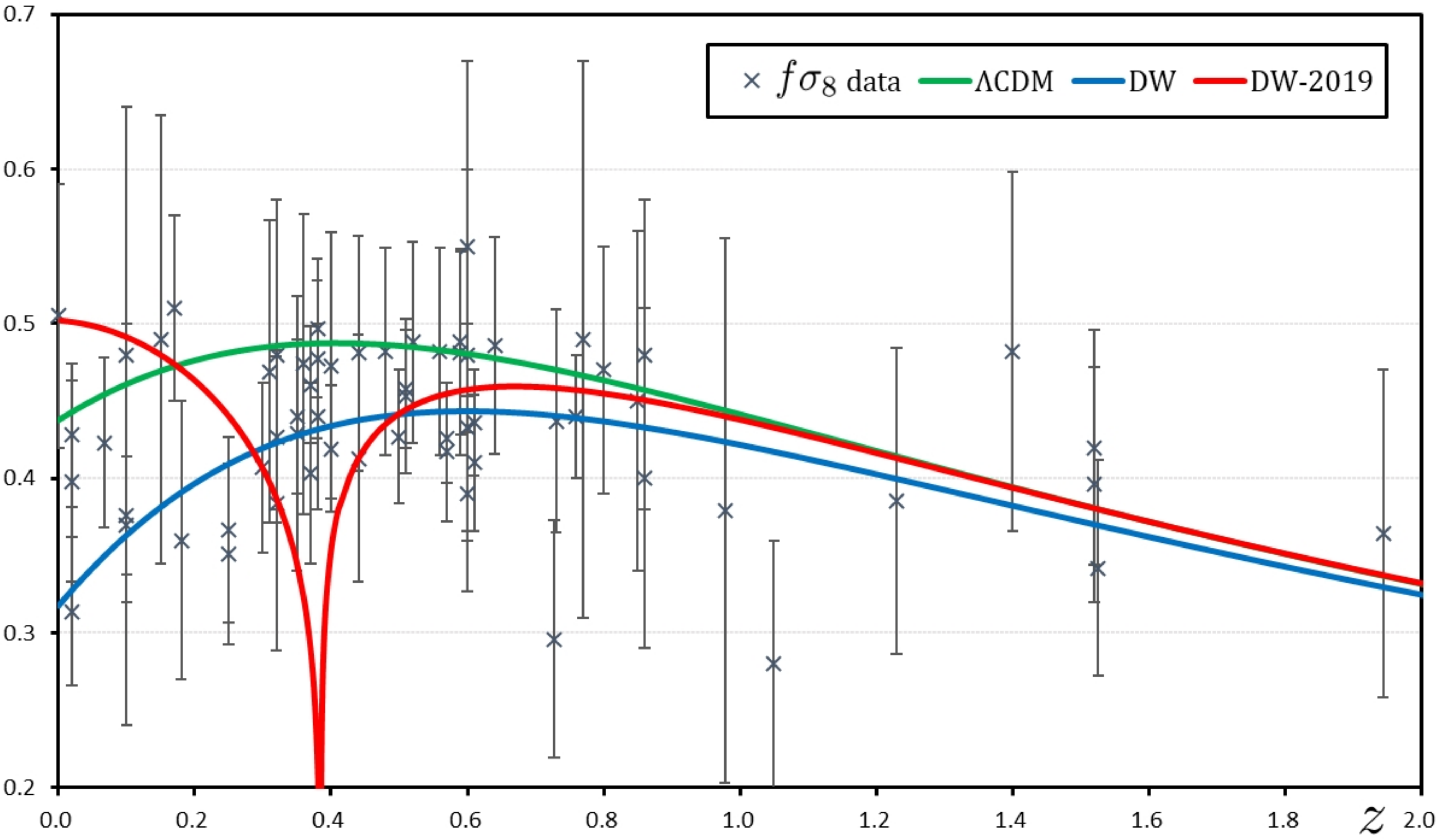}
\caption{\label{f_sigma8} The predicted $f\sigma_{8}$ at different redshift $z$ provided by $\Lambda$CDM, DW and DW-2019 models under the background\eqref{Hubble parameter for the background}, and the $f\sigma_{8}$ data from RSD measurements that are shown in TABLE \ref{TABLE RSD data} and \ref{TABLE RSD data 2} in Appendix.}
\end{figure}

 Using $N=-\ln(z+1)$, we plotted the numerical result of $f\sigma_{8}(z)$ under the DW-2019 model in FIG.\ref{f_sigma8}. As shown in FIG.\ref{f_sigma8}, our numerical result of $f\sigma_{8}(z)$ under the DW model is consistent with the result in \cite{nersisyan2017structure}. The predicted values of $f\sigma_{8}(z)$ under the DW-2019 model can not provide a good fit with the low-redshift RSD measurements. Moreover, there is a distinct and unnatural plummet of the growth rate curve at $z\simeq0.39$, which is caused by the above-mentioned discontinuity of the matter perturbation $\delta_{m}$ sourced by the strong nonlocal modification in the DW-2019 model. In another word, the DW-2019 model produces a strong nonlocal effect corresponding to the dark energy, which provides a more dramatic evolution of the matter perturbation $\delta_{m}$. In Sec.~\ref{sec:Gravitational Slip Equation}, based on the gravitational slip we will discuss it further. In addition, the clear difference of the predicted $f\sigma8$ between these models appears in the low-redshift range, which illustrates that the low-redshift RSD measurement is an efficient tool to distinguish and test these models.

\section{Gravitational Slip}
\label{sec:Gravitational Slip Equation}

 Whereas GR predicts $\Psi=\Phi$ in the presence of nonrelativistic matter, the difference between the amplitudes of the Newtonian ($\Psi$) and longitudinal ($\Phi$) gravitational potentials, called gravitational slip~\cite{PhysRevD.77.103513,PhysRevD.89.063538}, usually can be understood as the effective anisotropic stress in modified gravity theories. The gravitational slip is a key variable in the characterization of the physical origin of the dark energy. For convenience, for the DW-2019 model we defined the gravitational slip as
 \begin{equation}
 \begin{aligned}
 \label{the definition of gravitational slip}
 \frac{\Psi-\Phi}{\Psi+\Phi}=\frac{2\bar{V}}{1+\bar{U}+f(\bar{Y})+6\bar{V}}.
 \end{aligned}
 \end{equation}

 Based on the zeroth-order part of the (00) component of field equations~\eqref{00-component conformal time} and Eqs.~\eqref{solution of psi}\eqref{solution of phi}, $\Psi$ and $\Phi$ can be written into
 \begin{equation}
 \begin{aligned}
 \label{Psi and Phi eqaution of C_psi and C_phi}
 &\Psi=-\frac{3}{2}\,\frac{\mathcal{H}^{2}}{k^{2}}\,\delta_{m}\cdot C_{\Psi},\\
 &\Phi=-\frac{3}{2}\,\frac{\mathcal{H}^{2}}{k^{2}}\,\delta_{m}\cdot C_{\Phi},\\
 \end{aligned}
 \end{equation}
 where $C_{\Psi}$ and $C_{\Phi}$ are the modified factors sourced by the DW-2019 model, given by
 \begin{equation}
 \label{C_Psi}
 \begin{aligned}
 &C_{\Psi}=\frac{1+\bar{U}+f(\bar{Y})+8\bar{V}}{(1+\bar{U}+f(\bar{Y}))(1+\bar{U}+f(\bar{Y})+6\bar{V})}\cdot\bigg\{ 1+\bar{U}+f(\bar{Y})+\partial_{N}\bar{U}+f^{(1)}(\bar{Y})\partial_{N}\bar{Y}\\
 &\indent\indent\quad+\frac{1}{6}\big[ \partial_{N}\bar{X}\partial_{N}\bar{U}+\partial_{N}\bar{Y}\partial_{N}\bar{V}+\bar{V}(\partial_{N}\bar{X})^2 \big] \bigg\},
 \end{aligned}
 \end{equation}
 \begin{equation}
 \begin{aligned}
 \label{C_Phi}
 &C_{\Phi}=\frac{1+\bar{U}+f(\bar{Y})+4\bar{V}}{(1+\bar{U}+f(\bar{Y}))(1+\bar{U}+f(\bar{Y})+6\bar{V})}\cdot\bigg\{ 1+\bar{U}+f(\bar{Y})+\partial_{N}\bar{U}+f^{(1)}(\bar{Y})\partial_{N}\bar{Y}\\
 &\indent\indent\quad+\frac{1}{6}\big[ \partial_{N}\bar{X}\partial_{N}\bar{U}+\partial_{N}\bar{Y}\partial_{N}\bar{V}+\bar{V}(\partial_{N}\bar{X})^2 \big] \bigg\}.
 \end{aligned}
 \end{equation}
 When $C_{\Psi}$ and $C_{\Psi}$ both degenerate to $1$, these two potentials will reduce to those of GR.
\par
\begin{figure}[h]
 \centering
 \includegraphics[width=0.6\textwidth]{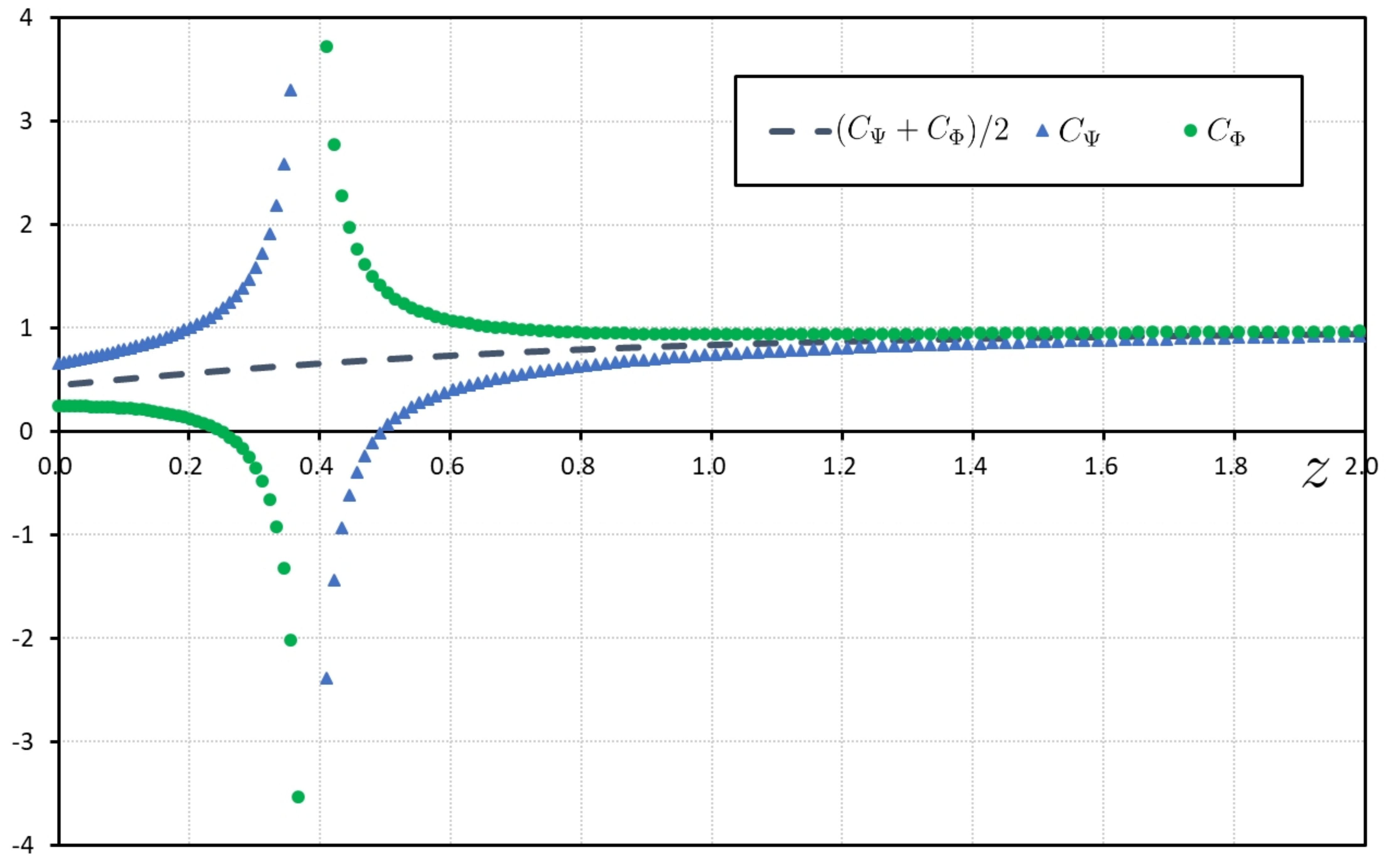}
 \caption{ \label{C_phi_C_psi} The evolving curves of $C_{\Psi}$, $C_{\phi}$ and $(C_{\Psi}+C_{\Phi})/2$. }
 \end{figure}

 With the basis of the reconstructed numerical results, we plotted the evolving curves of $C_{\Psi}$ and $C_{\Phi}$ in FIG.~\ref{C_phi_C_psi}, which illustrates that $C_{\Psi}$ and $C_{\Phi}$ both diverges at $z\simeq0.39$ where $1+\bar{U}+f(\bar{Y})+6\bar{V}=0$. On the contrary, $(C_{\Psi}+C_{\Phi})/2$ evolves smoothly without the divergency, which shows the sum of $\Psi$ and $\Phi$ always behaves regularly. This explains that in the sub-horizon limit ($k\gg\mathcal{H}$) the results from the linear perturbation theory is still reliable and self-consistent as a whole.
\par
\begin{figure}[h]
 \centering
 \includegraphics[width=0.6\textwidth]{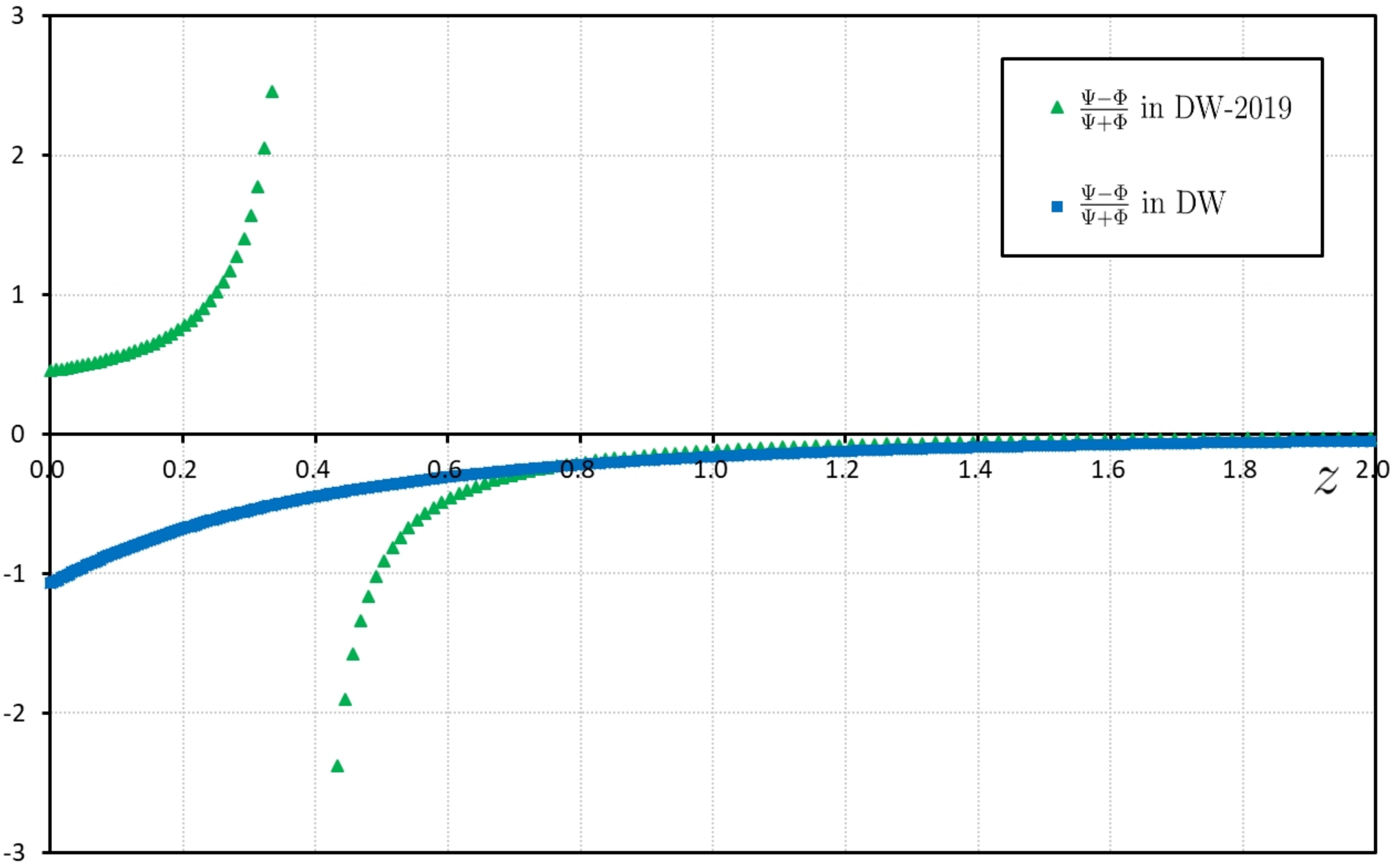}
 \caption{\label{Gravitataional slip} The evolution of the gravitational slip respectively in the DW and DW-2019 model.}
 \end{figure}

 On the other hand, the gravitational slip represents the anisotropic effect in the modified gravities, so the strong gravitational slip may influence the evolution of the matter perturbation $\delta_{m}$. FIG.~\ref{Gravitataional slip} shows the amplitude of the gravitational slip of the DW-2019 model is stronger than that of the DW model, which shows the DW-2019 model may provide more dark energy parts to impact the evolution of the matter perturbation $\delta_{m}$. And that is one possible cause of leading to the discontinuity of the growth rate $f\sigma_{8}$.

 \section{The Free Screening Mechanism}
 \label{sec:The Free Screening Mechanism}

 The time variation of the effective Newtonian gravitational constant is another important criterion to test modified gravities~\cite{belgacem2019testing,tian2019newtonian}. In the DW-2019 model, the effective Newtonian gravitational constant $G_{eff}=\big[1+\bar{U}+f(\bar{Y})\big]^{-1}\,G$ and its time variation is
 \begin{equation}
 \label{the basic time variation in DW-2019 model}
 \frac{\dot{G}_{eff}}{G_{eff}}=-\frac{ \partial_{N}\bar{U}+f^{(1)}(\bar{Y})\,\partial_{N}\bar{Y}
 }{1+\bar{U}+f(\bar{Y})}\,H.
 \end{equation}
 Using the numerical results, the time variation of Newtonian gravitational constant in DW-2019 model is given by
 \begin{equation}
 \begin{aligned}
 \label{dot(Geff)/Geff}
 \bigg|\frac{\dot{G}_{eff}}{G_{eff}}\bigg|\simeq0.717 H_{0}\sim \mathcal{O}(H_{0}).
 \end{aligned}
 \end{equation}
 On the other hand, Lunar Laser Ranging observation provides a strict limit on the time variation of Newtonian gravitational constant~\cite{belgacem2019testing,hofmann2018relativistic}
 \begin{equation}
 \begin{aligned}
 \label{the Lunar Laser Ranging limit}
 \frac{\dot{G}}{G}&=(7.1\pm7.6)\times10^{-14}\, yr^{-1}\\
 &=(0.99\pm1.06)\cdot(\frac{0.7}{h_{0}})\times10^{-3}\cdot H_{0}\\
 &\sim\mathcal{O}(10^{-3}\,H_{0}).
 \end{aligned}
 \end{equation}
 Hence DW-2019 model seems to be ruled out by Lunar Laser Ranging observation. However, there may exist a natural screening mechanism provided by the inverse scalar d'Alembertian. There is no reason to apply the FLRW solution in the strongly bound matter regime because the uneven matter distribution must curve spacetime. In order to solve this problem, connecting the cosmological regime with the strongly bound matter regime, one applied the McVittie metric~\cite{belgacem2019testing} as the background metric,
 \begin{equation}
 \begin{aligned}
 \label{the McVittie metric}
 d s^{2}=-\big[1-\Upsilon(r)-r^{2}H^{2}\big]dt^{2}-\frac{2r H}{\sqrt{1-\Upsilon(r)}}dr dt+\frac{1}{1-\Upsilon(r)}dr^{2}+r^{2}d\Omega^{2},
 \end{aligned}
 \end{equation}
 where $\Upsilon(r)=r_{s}/r$. $r_s$ is the Schwarzschild radius of the central object of mass M. The McVittie metric reduces to the FRW solution as $M\rightarrow0$, and to the Schwarzschild solution as $H\rightarrow0$. In order to test the modified gravities by the Lunar Laser Ranging, we should consider the circumstance of the Earth-Moon scales. For an actual observer , the central object can be simply regarded as Earth whose Schwarzschild radius is far less than its radius, which states $\Upsilon\ll1$. In fact, this condition is always reasonable as long as the central object is not a black hole. In addition, for the different cases of gravitationally bound systems, the general McVittie metric was presented by generalizing the form of $\Upsilon(r)$~\cite{belgacem2019testing}.
\par
 Based on the McVittie metric, one can obtain the Ricci scalar
 \begin{equation}
 \begin{aligned}
 \label{The Ricci scalar in McVittie metric}
 R(t,r)=12H^{2}+\frac{6\dot{H}}{\sqrt{1-\Upsilon}}+\frac{2}{r^{2}}\,\Upsilon+\frac{4}{r}\,\partial_{r}\Upsilon
 +\partial^{2}_{r}\Upsilon
 \end{aligned}
 \end{equation}
 and the scalar equations $S(t,r)=\Box J(t,r)$ can be expanded as
 \begin{equation}
 \begin{aligned}
 \label{S(t,r)=box J(t,r)}
 S(t,r)=&-\frac{1}{1-\Upsilon}\,\partial^{2}_{t}J-\frac{3H}{\sqrt{1-\Upsilon}}\,\partial_{t}J-\frac{1}{2}\frac{r\,\partial_{r}\Upsilon H}{(1-\Upsilon)^{3/2}}\,\partial_{t}J\\
 &+(1-\Upsilon-r^{2}H^{2})\,\partial^{2}_{r}J+\frac{2}{r}\,[1-\Upsilon-2r^{2}H^{2}-\frac{1}{2}r\,\partial_{r}\Upsilon]\,\partial_{r}J\\
 &-\frac{r\dot{H}}{\sqrt{1-\Upsilon}}\partial_{r}J
 -\frac{rH}{\sqrt{1-\Upsilon}}(\partial_{t}\partial_{r}J+\partial_{r}\partial_{t}J),
 \end{aligned}
 \end{equation}
 where $S$ represents the source of the nonlocal modification $J$ and $J=X,\,Y,\,V,\,U$.
 Applying the above condition of $\Upsilon\ll1$, the Ricci scalar and the scalar equations can be simplified into
 \begin{equation}
 \begin{aligned}
 \label{The Ricci scalar on Earth}
 R(t,r)=12H^{2}+6\dot{H}+\frac{2}{r^{2}}\,\Upsilon+\frac{4}{r}\,\partial_{r}\Upsilon+\partial^{2}_{r}\Upsilon,
 \end{aligned}
 \end{equation}
 \begin{equation}
 \begin{aligned}
 \label{S(t,r)=box J(t,r) on Earth}
 S(t,r)&=-\frac{1}{a^{3}}\,\partial_{t}(\,a^{3}\,\partial_{t}J\,)+\frac{1}{r^{2}}\,\partial_{r}(\,r^{2}\,\partial_{r}J\,)-r^{2}\,H^{2}\,\big[\,\partial^{2}_{r}J+\frac{1}{r}\,(4-\epsilon)\,\partial_{r}J\,\big]\\
 &\indent\indent-r\,H\,(\,\partial_{t}\partial_{r}J+\partial_{r}\partial_{t}J\,),
 \end{aligned}
 \end{equation}
 where $\epsilon\equiv-\frac{\dot{H}}{H^{2}}$ that is the slowly-varying Hubble parameter.
\par
 For the gravitationally bound systems with Earth-Moon scales, we have the constraint of $rH\ll1$ and the expression of $S(t,r)$ can be further simplified into
 \begin{equation}
 \begin{aligned}
 \label{S(t,r)=box J(t,r) in small scale}
 S(t,r)=-\frac{1}{a^{3}}\partial_{t}(a^{3}\partial_{t}J)+\frac{1}{r^{2}}\partial_{r}(r^{2}\partial_{r}J).
 \end{aligned}
 \end{equation}
\par
 For the scalar field $X$ ($J=X$), the source is the Ricci scalar ($S=R$). Because the source $R$ has the superimposed form of $R(t,r)=R_{cosmo}(t)+R_{static}(r)$ as shown in Eq.\eqref{The Ricci scalar on Earth}, $X$ can be chosen as a simple superimposed form $X(t,r)=X_{cosmo}(t)+X_{static}(r)$. It is worth mentioning that $R_{cosmo}(t)$ is provided by the expansion of the unverse which belongs to the large-scale effect and $R_{static}(r)$ is provided by the inhomogeneity of matter distribution which belongs to the small-scale effect.
\par
 For the case of the scalar field $Y$ whose source is $S=g^{\mu\nu}\partial_{\mu}X\partial_{\nu}X$, the scalar equation of $Y$ can be expanded as
 \begin{equation}
 \begin{aligned}
 \label{Box Y in small scale}
 -(\partial_{t}X_{cosmo})^{2}+(\partial_{r}X_{static})^{2}\simeq-\frac{1}{a^{3}}\partial_{t}(a^{3}\partial_{t}Y)+\frac{1}{r^{2}}\partial_{r}(r^{2}\partial_{r}Y).
 \end{aligned}
 \end{equation}
 Then $Y(t,r)$ has the superimposed form of $Y_{cosmo}(t)+Y_{static}(r)$ as well. Therefore, for the scalar fields $X$ and $Y$, the contributions from the large-scale effect and the small-scale effect are linearly superimposed.
\par
 However, for the scalar fields $V(t,r)$ and $U(t,r)$, things are going to change. Assuming that the nonlocal distortion function $f(Y)$~\eqref{f(Y)_fit} is still applicative in the small-scale range, the source function of $V$ is
 \begin{equation}
 \begin{aligned}
 \label{source of V}
 R\,f^{(1)}(Y)&=[R_{cosmo}(t)+R_{static}(r)]\cdot f^{(1)}(Y_{cosmo}(t)+Y_{static}(r))\\
 &\propto\, R_{cosmo}(t)\cdot f^{(1)}(Y_{cosmo}(t))\cdot f^{(1)}(Y_{static}(r))\\
 &\indent\quad+R_{static}(r)\cdot f^{(1)}(Y_{cosmo}(t))\cdot f^{(1)}(Y_{static}(r)).
 \end{aligned}
 \end{equation}
 Obviously the scalar field $V$ can be not written into the superimposed form like $X$ and $Y$, because its source function has the cross terms of time and space. From Eq.\eqref{Box U}, we can estimate that $U$ has an intricate form of $(t,r)$ including the cross terms of $(t,r)$ as well.
\par
 For the sake of illustration, considering the small-scale effect, we obtained the generalization of the time variation of Newtonian's constant qualitatively, given by
 \begin{equation}
 \begin{aligned}
 \label{f(1) Y_fit}
 \frac{\dot{G}_{eff}}{G_{eff}}\sim-\frac{\partial_{N}U(t,r)+2.153\,\boldsymbol{e}^{2.153\,(Y_{cosmo}+Y_{static}-16.97)}\,\partial_{N}Y_{cosmo}}
 {1+U(t,r)+\boldsymbol{e}^{2.153\,(Y_{cosmo}+Y_{static}-16.97)}}\cdot H.
 \end{aligned}
 \end{equation}
 Unlike the linear superposition of the large-scale effect and the small-scale effect in $X$ and $Y$, these two kinds of effects appear as the nonlinear recombination in the scalar fields $U(t,r)$ and $V(t,r)$. Therefore, there is the possibility that the small-scale effect is magnified in $V$ or $U$ to produce a free screening mechanism. With the help of the general McVittie metric, the deeper research can be studied in the future.

\section{Conclusions}

 In this work, we derive the first-order field equations of the DW-2019 model by the cosmological perturbation theory. In order to study the growth rate beyond the $\Lambda$CDM background, firstly we apply the reconstructing technique to obtain the evolution of the background fields in the DW-2019 model, and the nonlocal distortion function $f(Y)$ is fitted as $f(Y)\simeq\boldsymbol{e}^{2.153(Y-16.97)}$. For the purpose of testing the reasonability of our numerical method, we calculate the growth rate $f\sigma_8$ predicted by the DW model that has a good consistency with the result in \cite{nersisyan2017structure}, which shows our numerical method is feasible. In the DW-2019 model, based on the numerical results from the reconstructing process, the predicted growth rate $f\sigma_8(z)$ is obtained, which deviates from $f\sigma_8$ data of the RSD measurements to some extent, as shown in FIG.\ref{f_sigma8}. Even so, the DW-2019 model still can not be ruled out by the RSD measurements and its reliability should be tested further by the low-redshift RSD measurements. Moreover, for the DW-2019 model, the evolving curve of the growth rate $f\sigma_{8}$ has an unnatural plummet at $z\simeq0.39$. Based on the numerical results, we find the direct reason is $1+\bar{U}+f(\bar{Y})+6\bar{V}=0$ at $z\simeq0.39$, which leads to the divergency of $G_{N}$ in the second-order differential equation of $\delta_{m}$. From the perspective of theoretical analysis, the possible cause is that the DW-2019 model produces a strong nonlocal effect which behaves as the strong anisotropic stress corresponding to the dark energy in the low-redshift range to impact the evolution of the matter perturbation. At last, by the qualitative analysis of the DW-2019 model, we pointed out that the spacial dependence of the nonlocal modification in the small-scale range may provide a free screening mechanism, in the meantime, perhaps it will produce the correction for the matter perturbation in the low-redshift range as well.

\acknowledgments
We are grateful to the referees for valuable comments. This work was supported by the National Natural Science Foundation of China~(Grant No.11571342).


\newpage
\appendix
\section*{Appendix}
\label{Appendix}

\begin{table}[h]
\centering
\caption{\label{TABLE RSD data} The $f\,\sigma_{8}$ data provided by the RSD measurements from various sources\cite{kazantzidis2018evolution}.}
\begin{tabular}{p{4cm}p{2cm}p{4cm}p{1.5cm}p{2cm}}
\hline
Survey & $z$ & $f\,\sigma_{8}$  & Ref. & Year \\ \hline
\hline
\footnotesize SDSS-LRG                        &\footnotesize  0.35                   &\footnotesize  0.44 $\pm$ 0.05         & \footnotesize \cite{song2009reconstructing}                &\footnotesize 2006       \\
\footnotesize VVDS                            &\footnotesize  0.77                   &\footnotesize  0.49 $\pm$ 0.18         & \footnotesize \cite{song2009reconstructing}                &\footnotesize 2009       \\
\footnotesize 2dFGRS                          &\footnotesize  0.17                   &\footnotesize  0.51 $\pm$ 0.06         & \footnotesize \cite{song2009reconstructing}                &\footnotesize 2009       \\
\footnotesize 2MRS                            &\footnotesize  0.02                   &\footnotesize  0.314 $\pm$ 0.048       & \footnotesize \cite{davis2011local},\cite{hudson2012growth}       &\footnotesize 2010       \\
\footnotesize SnIa-IRAS                       &\footnotesize  0.02                   &\footnotesize  0.398 $\pm$ 0.065       & \footnotesize \cite{hudson2012growth},\cite{turnbull2012cosmic}   &\footnotesize 2011       \\
\footnotesize SDSS-LRG-200                    &\footnotesize  0.25                   &\footnotesize  0.3512 $\pm$ 0.00583    & \footnotesize \cite{samushia2012interpreting}              &\footnotesize 2011       \\
\footnotesize SDSS-LRG-200                    &\footnotesize  0.37                   &\footnotesize  0.4602 $\pm$ 0.0378     & \footnotesize \cite{samushia2012interpreting}              &\footnotesize 2011       \\
\footnotesize SDSS-LRG-60                     &\footnotesize  0.25                   &\footnotesize  0.3665 $\pm$ 0.0601     & \footnotesize \cite{samushia2012interpreting}              &\footnotesize 2011       \\
\footnotesize SDSS-LRG-60                     &\footnotesize  0.37                   &\footnotesize  0.4031 $\pm$ 0.0586     & \footnotesize \cite{samushia2012interpreting}              &\footnotesize 2011       \\
\footnotesize WiggleZ                         &\footnotesize  0.44                   &\footnotesize  0.413 $\pm$ 0.08        & \footnotesize \cite{blake2012wigglez}                      &\footnotesize 2012       \\
\footnotesize WiggleZ                         &\footnotesize  0.6                    &\footnotesize  0.39 $\pm$ 0.063        & \footnotesize \cite{blake2012wigglez}                      &\footnotesize 2012       \\
\footnotesize WiggleZ                         &\footnotesize  0.73                   &\footnotesize  0.437 $\pm$ 0.072       & \footnotesize \cite{blake2012wigglez}           &\footnotesize 2012       \\
\footnotesize 6dFGS                           &\footnotesize  0.067                  &\footnotesize  0.423 $\pm$ 0.055       & \footnotesize \cite{beutler20126df}             &\footnotesize 2012       \\
\footnotesize SDSS-BOSS                       &\footnotesize  0.3                    &\footnotesize  0.407 $\pm$ 0.055       & \footnotesize \cite{tojeiro2012clustering}      &\footnotesize 2012       \\
\footnotesize SDSS-BOSS                       &\footnotesize  0.4                    &\footnotesize  0.419 $\pm$ 0.041       & \footnotesize \cite{tojeiro2012clustering}      &\footnotesize 2012       \\
\footnotesize SDSS-BOSS                       &\footnotesize  0.5                    &\footnotesize  0.427 $\pm$ 0.043       & \footnotesize \cite{tojeiro2012clustering}      &\footnotesize 2012       \\
\footnotesize SDSS-BOSS                       &\footnotesize  0.6                    &\footnotesize  0.433 $\pm$ 0.067       & \footnotesize \cite{tojeiro2012clustering}      &\footnotesize 2012       \\
\footnotesize Vipers                          &\footnotesize  0.8                    &\footnotesize  0.47 $\pm$ 0.08         & \footnotesize \cite{de2013vimos}                &\footnotesize 2013       \\
\footnotesize SDSS-DR7-LRG                    &\footnotesize  0.35                   &\footnotesize  0.429 $\pm$ 0.089       & \footnotesize \cite{chuang2013modelling}        &\footnotesize 2013       \\
\footnotesize GAMA                            &\footnotesize  0.18                   &\footnotesize  0.36 $\pm$ 0.09         & \footnotesize \cite{blake2013galaxy}            &\footnotesize 2013       \\
\footnotesize GAMA                            &\footnotesize  0.38                   &\footnotesize  0.44 $\pm$ 0.06         & \footnotesize \cite{blake2013galaxy}            &\footnotesize 2013       \\
\footnotesize BOSS-LOWZ                       &\footnotesize  0.32                   &\footnotesize  0.384 $\pm$ 0.095       & \footnotesize \cite{sanchez2014clustering}      &\footnotesize 2013       \\
\footnotesize SDSS DR10/11                    &\footnotesize  0.32                   &\footnotesize  0.48 $\pm$ 0.1          & \footnotesize \cite{sanchez2014clustering}      &\footnotesize 2013       \\
\footnotesize SDSS DR10/11                    &\footnotesize  0.57                   &\footnotesize  0.417 $\pm$ 0.045       & \footnotesize \cite{sanchez2014clustering}      &\footnotesize 2013       \\
\footnotesize SDSS-MGS                        &\footnotesize  0.15                   &\footnotesize  0.49 $\pm$ 0.145        & \footnotesize \cite{howlett2015clustering}      &\footnotesize 2015       \\
\footnotesize SDSS-veloc                      &\footnotesize  0.1                    &\footnotesize  0.37 $\pm$ 0.13         & \footnotesize \cite{feix2015growth}             &\footnotesize 2015       \\
\footnotesize FastSound                       &\footnotesize  1.4                    &\footnotesize  0.482 $\pm$ 0.116       & \footnotesize \cite{okumura2016subaru}          &\footnotesize 2015       \\
\footnotesize SDSS-CMASS                      &\footnotesize  0.59                   &\footnotesize  0.488 $\pm$ 0.06        & \footnotesize \cite{chuang2016clustering}       &\footnotesize 2016       \\
\footnotesize BOSS DR12                       &\footnotesize  0.38                   &\footnotesize  0.497 $\pm$ 0.045       & \footnotesize \cite{alam2017clustering}         &\footnotesize 2016       \\
\footnotesize BOSS DR12                       &\footnotesize  0.51                   &\footnotesize  0.458 $\pm$ 0.038       & \footnotesize \cite{alam2017clustering}         &\footnotesize 2016       \\
\footnotesize BOSS DR12                       &\footnotesize  0.61                   &\footnotesize  0.436 $\pm$ 0.034       & \footnotesize \cite{alam2017clustering}         &\footnotesize 2016       \\ \hline
\end{tabular}
\end{table}

\begin{table}[bp]
\centering
\caption{\label{TABLE RSD data 2} The $f\,\sigma_{8}$ data following the above table}
\begin{tabular}{p{4cm}p{2cm}p{4cm}p{1.5cm}p{2cm}}
\hline
Survey & $z$ & $f\,\sigma_{8}$  & Ref. & Year \\ \hline
\hline
\footnotesize BOSS DR12                       &\footnotesize  0.38                   &\footnotesize  0.477 $\pm$ 0.051       & \footnotesize \cite{beutler2016clustering}      &\footnotesize 2016       \\
\footnotesize BOSS DR12                       &\footnotesize  0.51                   &\footnotesize  0.453 $\pm$ 0.05        & \footnotesize \cite{beutler2016clustering}      &\footnotesize 2016       \\
\footnotesize BOSS DR12                       &\footnotesize  0.61                   &\footnotesize  0.41 $\pm$ 0.044        & \footnotesize \cite{beutler2016clustering}      &\footnotesize 2016       \\
\footnotesize Vipers v7                       &\footnotesize  0.76                   &\footnotesize  0.44 $\pm$ 0.04         & \footnotesize \cite{wilson2016geometric}        &\footnotesize 2016       \\
\footnotesize Vipers v7                       &\footnotesize  1.05                   &\footnotesize  0.28 $\pm$ 0.08         & \footnotesize \cite{wilson2016geometric}        &\footnotesize 2016       \\
\footnotesize BOSS LOWZ                       &\footnotesize  0.32                   &\footnotesize  0.427 $\pm$ 0.056       & \footnotesize \cite{gil2016clustering}          &\footnotesize 2016       \\
\footnotesize BOSS CMASS                      &\footnotesize  0.57                   &\footnotesize  0.426 $\pm$ 0.029       & \footnotesize \cite{gil2016clustering}          &\footnotesize 2016       \\
\footnotesize Vipers                          &\footnotesize  0.727                  &\footnotesize  0.296 $\pm$ 0.0765      & \footnotesize \cite{hawken2017vimos}            &\footnotesize 2016       \\
\footnotesize 6dFGS+SnIa                      &\footnotesize  0.02                   &\footnotesize  0.428 $\pm$ 0.0465      & \footnotesize \cite{huterer2017testing}         &\footnotesize 2016       \\
\footnotesize Vipers                          &\footnotesize  0.6                    &\footnotesize  0.48 $\pm$ 0.12         & \footnotesize \cite{de2017vimos}                &\footnotesize 2016       \\
\footnotesize Vipers                          &\footnotesize  0.86                   &\footnotesize  0.48 $\pm$ 0.1          & \footnotesize \cite{de2017vimos}                &\footnotesize 2016       \\
\footnotesize Vipers PDR-2                    &\footnotesize  0.6                    &\footnotesize  0.55 $\pm$ 0.12         & \footnotesize \cite{pezzotta2017vimos}          &\footnotesize 2016       \\
\footnotesize Vipers PDR-2                    &\footnotesize  0.86                   &\footnotesize  0.4 $\pm$ 0.11          & \footnotesize \cite{pezzotta2017vimos}          &\footnotesize 2016       \\
\footnotesize SDSS DR13                       &\footnotesize  0.1                    &\footnotesize  0.48 $\pm$ 0.16         & \footnotesize \cite{feix2017speed}              &\footnotesize 2016       \\
\footnotesize 2MTF                            &\footnotesize  0.001                  &\footnotesize  0.505 $\pm$ 0.085       & \footnotesize \cite{howlett20172mtf}            &\footnotesize 2017       \\
\footnotesize Vipers PDR-2                    &\footnotesize  0.85                   &\footnotesize  0.45 $\pm$ 0.11         & \footnotesize \cite{mohammad2018vimos}          &\footnotesize 2017       \\
\footnotesize BOSS DR12                       &\footnotesize  0.31                   &\footnotesize  0.469 $\pm$ 0.098       & \footnotesize \cite{wang2018clustering}         &\footnotesize 2017       \\
\footnotesize BOSS DR12                       &\footnotesize  0.36                   &\footnotesize  0.474 $\pm$ 0.097       & \footnotesize \cite{wang2018clustering}         &\footnotesize 2017       \\
\footnotesize BOSS DR12                       &\footnotesize  0.4                    &\footnotesize  0.473 $\pm$ 0.086       & \footnotesize \cite{wang2018clustering}         &\footnotesize 2017       \\
\footnotesize BOSS DR12                       &\footnotesize  0.44                   &\footnotesize  0.481 $\pm$ 0.076       & \footnotesize\cite{wang2018clustering}        &\footnotesize 2017       \\
\footnotesize BOSS DR12                       &\footnotesize  0.48                   &\footnotesize  0.482 $\pm$ 0.067       & \footnotesize \cite{wang2018clustering}         &\footnotesize 2017       \\
\footnotesize BOSS DR12                       &\footnotesize  0.52                   &\footnotesize  0.488 $\pm$ 0.065       & \footnotesize \cite{wang2018clustering}         &\footnotesize 2017       \\
\footnotesize BOSS DR12                       &\footnotesize  0.56                   &\footnotesize  0.482 $\pm$ 0.067       & \footnotesize \cite{wang2018clustering}         &\footnotesize 2017       \\
\footnotesize BOSS DR12                       &\footnotesize  0.59                   &\footnotesize  0.481 $\pm$ 0.066       & \footnotesize \cite{wang2018clustering}         &\footnotesize 2017       \\
\footnotesize BOSS DR12                       &\footnotesize  0.64                   &\footnotesize  0.486 $\pm$ 0.07        & \footnotesize \cite{wang2018clustering}         &\footnotesize 2017       \\
\footnotesize SDSS DR7                        &\footnotesize  0.1                    &\footnotesize  0.376 $\pm$ 0.038       & \footnotesize \cite{shi2018mapping}             &\footnotesize 2017       \\
\footnotesize SDSS-IV                         &\footnotesize  1.52                   &\footnotesize  0.42 $\pm$ 0.076        & \footnotesize \cite{gil2018clustering}          &\footnotesize 2018       \\
\footnotesize SDSS-IV                         &\footnotesize  1.52                   &\footnotesize  0.396 $\pm$ 0.076       & \footnotesize \cite{hou2018clustering}          &\footnotesize 2018       \\
\footnotesize SDSS-IV                         &\footnotesize  0.978                  &\footnotesize  0.379 $\pm$ 0.176       & \footnotesize \cite{zhao2018clustering}         &\footnotesize 2018       \\
\footnotesize SDSS-IV                         &\footnotesize  1.23                   &\footnotesize  0.385 $\pm$ 0.099       & \footnotesize \cite{zhao2018clustering}         &\footnotesize 2018       \\
\footnotesize SDSS-IV                         &\footnotesize  1.526                  &\footnotesize  0.342 $\pm$ 0.07        & \footnotesize \cite{zhao2018clustering}         &\footnotesize 2018       \\
\footnotesize SDSS-IV                         &\footnotesize  1.944                  &\footnotesize  0.364 $\pm$ 0.106       & \footnotesize \cite{zhao2018clustering}         &\footnotesize 2018       \\ \hline
\end{tabular}
\end{table}

\end{document}